\documentclass[reqno,12pt]{article}
\usepackage{amsmath,amsfonts,amssymb,amsthm,amstext,amscd}
\usepackage{hyperref}
\usepackage{epsfig}
\usepackage{graphicx}
\usepackage{caption}
\usepackage{subcaption}
\usepackage{comment}
\usepackage{graphicx}
\usepackage{cite}
\makeatletter \@addtoreset{equation}{section}
\usepackage{latexsym}

\usepackage{enumerate}
\usepackage{color}
\usepackage{setspace}
\usepackage{float}
\usepackage[mathscr]{eucal}

\makeatletter\renewcommand\section{\@startsection {section}{1}{\z@}%
                                                                                                                                                                                                                                                                                                                                                                                                                                                                                                                                                                                                                                                                                                                                                                                                                                                                                                                                                                                                                                                                                                                                                                                                                                                                                                                                                                                                                                                                                                                                                                                                                                                                                                                                                                                                                                                                                                   {-3.5ex \@plus -1ex \@minus -.2ex}
                                                                                                                                                                                                                                                                                                                                                                                                                                                                                                                                                                                                                                                                                                                                                                                                                                                                                                                                                                                                                                                                                                                                                                                                                                                                                                                                                                                                                                                                                                                                                                                                                                                                                                                                                                                                                                                                                                   {2.3ex \@plus.2ex}%
                                                                                                                                                                                                                                                                                                                                                                                                                                                                                                                                                                                                                                                                                                                                                                                                                                                                                                                                                                                                                                                                                                                                                                                                                                                                                                                                                                                                                                                                                                                                                                                                                                                                                                                                                                                                                                                                                                   {\normalfont\large\bfseries}}
\renewcommand\subsection{\@startsection{subsection}{2}{\z@}%
                                                                                                                                                                                                                                                                                                                                                                                                                                                                                                                                                                                                                                                                                                                                                                                                                                                                                                                                                                                                                                                                                                                                                                                                                                                                                                                                                                                                                                                                                                                                                                                                                                                                                                                                                                                                                                                                                                                                                                                                                                                     {-3.25ex\@plus -1ex \@minus -.2ex}%
                                                                                                                                                                                                                                                                                                                                                                                                                                                                                                                                                                                                                                                                                                                                                                                                                                                                                                                                                                                                                                                                                                                                                                                                                                                                                                                                                                                                                                                                                                                                                                                                                                                                                                                                                                                                                                                                                                                                                                                                                                                     {1.5ex \@plus .2ex}%
                                                                                                                                                                                                                                                                                                                                                                                                                                                                                                                                                                                                                                                                                                                                                                                                                                                                                                                                                                                                                                                                                                                                                                                                                                                                                                                                                                                                                                                                                                                                                                                                                                                                                                                                                                                                                                                                                                                                                                                                                                                     {\normalfont\bfseries}}

\newcommand{\be}{\begin{equation}}
\newcommand{\ee}{\end{equation}}
\newcommand{\bea}{\begin{eqnarray}}
\newcommand{\eea}{\end{eqnarray}}
\newcommand{\bse}{\begin{subequations}}
\newcommand{\ese}{\end{subequations}}
\newcommand{\beqa}{\begin{eqnarray}}
\newcommand{\eeqa}{\end{eqnarray}}
\newcommand{\beqar}{\begin{eqnarray*}}
\newcommand{\eeqar}{\end{eqnarray*}}
\newcommand{\bi}{\begin{itemize}}
\newcommand{\ei}{\end{itemize}}
\newcommand{\bn}{\begin{enumerate}}
\newcommand{\en}{\end{enumerate}}

\newcommand{\ba}{\begin{array}}
\newcommand{\ea}{\end{array}}
\newcommand{\bc}{\begin{center}}
\newcommand{\ec}{\end{center}}

\newcommand{\ie}{{\em i.e.}\ }
\newcommand{\eg}{{\em e.g.} }

\newcommand{\mrd}{\mathrm{d}}

\newcommand{\mrH}{\mathrm{H}}

\definecolor{mgreen}{rgb}{0,0.6,0}
\definecolor{darkred}{rgb}{0.7,0,0}

\definecolor{darkblue}{rgb}{0.2,0.2,0.6}
\definecolor{darkgreen}{rgb}{0.2,0.6,0.2}

\begin{document}

\begin{titlepage}

\begin{flushright}\vspace{-3cm}
{\small
IPM/P-2015/073 \\
December 18, 2015}\end{flushright}
\vspace{0.5cm}

\begin{center}
\centerline{{\Large{\bf{Solution Phase Space and   Conserved Charges}}}} \vspace{4mm}
\centerline{{\large{\bf{A General Formulation for Charges Associated with Exact Symmetries}}}}
 \vspace{9mm}

\centerline{\large{{{K. Hajian\footnote{e-mail: kamalhajian@ipm.ir} \ and \  M.M. Sheikh-Jabbari\footnote{e-mail:
jabbari@theory.ipm.ac.ir}}}}}

\vspace{3mm}
\normalsize
 \textit{School of Physics, Institute for Research in Fundamental
Sciences (IPM), \\ P.O. Box 19395-5531, Tehran, Iran}
\vspace{5mm}

\begin{abstract}
\noindent
We  provide a general formulation for calculating conserved charges for solutions to generally covariant gravitational theories with possibly other internal gauge symmetries, in any dimensions and with generic asymptotic behaviors. These solutions are generically specified by a number of exact (continuous, global) symmetries and some parameters. We define ``parametric variations" as field perturbations generated by variations of the solution parameters. Employing the covariant phase space method, we  establish that the set of these solutions (up to pure gauge transformations) form a phase space, the \emph{solution phase space}, and that the tangent space of this phase space includes the parametric variations. We then compute  conserved charge variations associated with the exact symmetries of the family of solutions, caused by parametric variations.  Integrating the charge variations over a path in the solution phase space, we define the conserved charges.  In particular, we revisit ``black hole entropy as a conserved charge'' and the derivation of the first law of black hole thermodynamics. We show that the solution phase space setting enables us to define black hole entropy by an integration over any compact, codminesion-2, smooth spacelike surface encircling the hole, as well as to a natural generalization of Wald and Iyer-Wald analysis to cases involving gauge fields.

\end{abstract}


\end{center}

\end{titlepage}
\setcounter{footnote}{0}
\renewcommand{\baselinestretch}{1.05}  

\addtocontents{toc}{\protect\setcounter{tocdepth}{2}}
\tableofcontents



\section{Introduction}

Since the seminal theorems by Emmy Noether, the notion of conserved charges has been linked to the concept of symmetry. Being a generally invariant theory and in lack of a globally defined time direction, defining conserved charges in a covariant way has been a challenge within General Relativity (GR). There have been many proposals since the early days of GR for computing conserved charges associated with \emph{exact} isometries of spacetime, the Killing vector fields, or the asymptotic isometries (\eg in asymptotic flat or anti-de Sitter spacetimes). The first covariant formula for conserved charges associated with Killing vectors has been due to Arthur Komar \cite{Komar:1958wp}, which was shortly followed by the Arnowitt-Deser-Misner (ADM) papers \cite{Arnowitt:1959ah,Arnowitt:1960es,Arnowitt:1962hi} which define charges associated with symmetries of asymptotic flat space on constant time slices, as well as papers by Bondi \emph{et al}. \cite{Bondi:1962px,Sachs:1962zza} which define the charges associated with asymptotic flat isometries at null infinity. These have also been extended to asymptotic anti-de Sitter (AdS) geometries \cite{Ashtekar:1984zz,Ashtekar:1999jx,Aros:1999id} (see Ref. \cite{Hollands:2005wt} for a nice overview on this case) and asymptotic flat solutions to higher derivative theories of gravity \cite{Deser:2002jk}. These methods have practical advantages, easy to work with, and are intuitive. Nonetheless, they have their own shortcomings. For example, they are not covariant enough or crucially depend on the form of Einstein-Hilbert action or on the asymptotic behavior of fields (\eg fit for asymptotic flat spacetimes). 

In a related line of developments, one may ask for a ``covariant'' Hamiltonian formulation of generally invariant theories. A key object in this formulation is the construction of phase space and its symplectic structure. To work out such a ``covariant phase space'' one may try to base the construction on the Lagrangian or action formulation. In fact this is easily possible, once we recall that the symplectic two-from may be read from the surface term in the variation of the action. As a simple illustrative example, let us consider a single particle Lagrangian $L=\dot q^2-V(q)$ and that $\delta L=(\ddot q-V')\delta q+\mrd(p\delta q)/\mrd t$. The second variation of the on-shell action will then yield $\mrd(\delta p\wedge\delta q)/\mrd t$, which is nothing but the time derivative of the symplectic structure two-form computed over the tangent space of the phase space parametrized by $\delta q, \delta p$; for example see Ref. \cite{Henneaux:1992ig}. Of course in extending this simple example  to field theories (which have an infinite-dimensional, continuous phase space), especially when there are gauge symmetries which can lead to degenerate directions on the symplectic two-form, one may face extra complications to tackle. This latter has been studied and analyzed systematically, \eg for gauge theories \cite{Henneaux:1992ig} or for gravity in series of precisely formulated papers by R. Wald and collaborators. This construction makes a direct relation with the analysis of conserved charges, once we recall the Noether theorem and that the charges are read from the same surface term. {The \emph{covariant phase space method} (CPSM), initiated in Refs. \cite{Lee:1990gr,Ashtekar:1987hia,Ashtekar:1990gc} and studied more in Refs. \cite{Barnich:2001jy,Barnich:2007bf}, is now a very well-developed topic (\eg see Refs. \cite{Szabados:2004vb,Compere:2007az,Hajian:2015eha} as reviews and references therein).} This method has been employed to define the conserved charges in a covariant way, and also in gauge field theories \eg see Refs. \cite{Ashtekar:1987hia,Ashtekar:1990gc, Wald:1999wa, Barnich:2007bf, Barnich:2003xg}.

In a different viewpoint general relativity like Maxwell or Yang-Mills gauge field theories has a local symmetry, a redundancy in the description: not all of the degrees of freedom encoded in dynamical fields are  associated with physical observables; physically distinct field configurations are those defined ``up to gauge transformations.'' These local gauge symmetries are not \emph{generically} a subject of Noether's theorems; \ie generically there are no conserved charges associated with the gauge transformations (see Ref. \cite{Avery:2015rga} for a recent discussion on this point.)

Let us consider a theory with gauge symmetries in which a generic solution to its field equations is denoted by $\Phi$. \emph{Exact symmetries} are the (nontrivial) subset of gauge transformations generated by $\eta$, for which $\delta_\eta\Phi=0$; \ie exact symmetries are the gauge transformations which do not move us in the solution space. Such $\eta$'s are hence in general field dependent, $\eta=\eta[\Phi]$. If we consider diffeomorphisms, these exact symmetries are the isometries of spacetime which are generated by Killing vector fields. In the presence of gauge fields, the exact symmetries are not limited to Killing vectors, and there could be a subset of internal gauge transformation which does not change a given solution. For example, the Maxwell gauge field corresponding to a set of static charges, in the static gauge, is invariant under gauge transformation generated by $\lambda={\rm const.}$, the global part of the $U(1)$ gauge transformations.   Moreover, in the presence of other gauge fields, the spacetime isometries may not necessarily keep the gauge field intact; they may transform the gauge field up to an internal gauge transformation and hence do not physically change the solution.  In this sense, it has been argued that the notion of ``invariance'' under exact symmetries should be extended to invariance up to internal gauge transformations. This latter will have very interesting consequences which have been considered in some previous works \cite{Barnich:2007bf,Sudarsky:1992ty, Sudarsky:1993kh, McCormick:2013nkb,HSS:2013lna, Prabhu:2015vua}.

Given any field theory, one may consider the set of solutions ${\cal F}$ with prescribed boundary and initial conditions  and  denote generic field configurations in ${\cal F}$, collectively by  $\Phi$. Each element in ${\cal F}$ may  be identified up to \emph{pure gauge transformations} (which we will  precisely define)  by some number of continuous real parameters $p_\alpha$. Hereafter, we will name the set of all given solutions spanned by parameters $p_\alpha$ the \emph{solution space}. That is, the parameters $p_\alpha$ may be viewed as ``coordinates'' on the solution space.\footnote{The solution may  also have some discrete parameters which will not appear in our discussions in this paper. Note also that the exact symmetry generators $\eta$ have implicit dependence on the parameters of the solution $p_\alpha$.} One can then consider field perturbations around a given solution $\delta\Phi$. In our setting in this paper, we consider only perturbations $\delta\Phi$ which satisfy linearized field equations (around $\Phi$). In a theory with local symmetries, the set of such field perturbations will, by definition, include two important subclasses: (1) those generated by gauge (local symmetry) transformations, which will be denoted by $\delta_\epsilon\Phi$ and (2) those generated by the variation of parameters in the solution space, the parametric variations, denoted by $\hat\delta\Phi$.\footnote{In general $\delta\Phi$ are not limited to these two classes. There are generically $\delta\Phi$ associated with waves corresponding to propagating degrees of freedom on the background $\Phi$. There are, however, interesting cases, like 3D gravity, 3D Chern-Simons theories or modes around near horizon extremal geometries where there are no propagating degrees of freedom. For these cases these two classes can be exhaustive.} In the generally covariant theories we are studying in this work, $\delta_\epsilon\Phi$ include field perturbations generated by infinitesimal diffeomorphisms $\delta x^\mu=-\epsilon^\mu(x)$ and hence $\delta_\epsilon\Phi=\mathscr{L}_\epsilon\Phi$, where $\mathscr{L}_\epsilon$ denotes the Lie derivative along vector field $\epsilon$. If we have other internal gauge symmetries, like in Einstein-Maxwell or Einstein-Yang-Mills theories, $\delta_\epsilon\Phi$ will also include such internal gauge transformations. Such field perturbations have been very much analyzed especially in the context of ``asymptotic symmetry groups.'' For the example of 3D gravity, 3D Chern-Simons theory, and near-horizon extremal geometries for some recent works see \eg Refs. \cite{Barnich:2014cwa,Compere:2014cna,Compere:2015knw,CHSS:2015mza,CHSS:2015bca} and references therein. Note that perturbations $\delta_\epsilon\Phi$ can be decomposed into two subclasses, those where $\delta_\epsilon\Phi=0$, which are nothing but the exact symmetries, and those where $\delta_\epsilon\Phi\neq 0$.

Our main goal in this paper is to analyze and establish the intimate connection of parametric variations $\hat\delta\Phi$ with the conserved charges associated with the {exact symmetries}. To this end, we employ the covariant phase space method and  pay special attention to the parametric variations $\hat\delta\Phi$. We show that the set of solutions  forms a well-defined phase space even when the tangent space to a given point associated with a solution $\Phi$  is restricted to $\hat\delta\Phi$. We will denote this phase space by ${\cal F}_p$. We then use the standard  covariant phase space method to read variation/perturbation due to $\hat\delta\Phi$ in conserved charges associated with exact symmetries. After discussing the integrability of these charge variations over ${\cal F}_p$, we integrate the charge variations over a path in the space of solutions to compute the charge associated with exact symmetries of each field configuration $\Phi$. 

Our method for computing charges associated with exact symmetries, and in particular focusing on the solution space,  brings many advantages: the charge variations are now computed by an integration over an arbitrary smooth, closed, and compact codimension-2 spacelike surface; we are not confined to taking the integrals at the asymptotic region of the spacetime. Of course, to completely specify the charges from charge variations, we need to choose a reference point in the solution space, for which as we will see, in each case we have a clearly preferred choice. Dealing with parametric perturbations and exact symmetries, our formulation is free of the ambiguities appearing in the usual Noether theorems
(see Refs. \cite{Iyer:1994ys,Wald:1999wa} for a thorough  analysis of such ambiguities).

The rest of this paper is organized as follows. In Sec. \ref{sec-2}, to make our analysis self-contained we provide a review of covariant phase space method, construction of the presymplectic current and the integrability condition for the charges. In Sec. \ref{sec-3}, we specialize discussions of Sec. \ref{sec-2} to the parametric variations. We construct the corresponding conserved charges associated with exact symmetries and the associated phase space. In Sec. \ref{sec-4}, to illustrate how our prescription for computing charges works, we carry out  the explicit computations  for two well-known solutions, the Kerr-AdS black hole and the near-horizon extremal Kerr-Newman geometry, and discuss interesting details in each example. In Sec. \ref{sec-5},
we use our machinery for computing the entropy variation and hence entropy in geometries admitting a Killing horizon. Our result here is essentially a repetition of Wald's seminal work \cite{Wald:1993nt}, with an important extension: the entropy need not be computed by an integral over the codimension-2 bifurcation surface of the bifurcate Killing horizon; it could be computed by integrating the corresponding density over any closed, compact spacelike codimension-2 surface. Similarly, other charges may be computed over any arbitrary surface. In this section we also revisit Iyer-Wald derivation \cite{Iyer:1994ys} of the first law of thermodynamics for charge perturbations of black holes and show how our formulation extends and generalize it in two different ways, especially when the gauge field charges are present. The last section is devoted to a summary and outlook. In some appendixes we have gathered some technical details of the analysis and calculations.

\section{Review of covariant phase space method and conserved charges}\label{sec-2}

The CPSM provides a systematic way of calculating variations or perturbations of conserved charges in generic theories with local gauge symmetries, in particular generally covariant theories. This method was primarily developed in the papers of Wald \emph{et al} \cite{Lee:1990gr,Wald:1993nt,Iyer:1994ys}.  Wald's approach, which is what we will review below, is based on action formulation. There is another way of formulating the CPSM, developed by Barnich \emph{et al}., based on equations of motion (instead of action) \cite{Barnich:2007bf,Barnich:2001jy}. In this work we discuss a less appreciated application of the CPSM for computing the conserved charges associated with exact symmetries. As we will see for this class (see also Ref. \cite{Barnich:2007bf}), the Wald \emph{et al.} and Barnich \emph{et al.} formulations become equivalent. We therefore briefly review  the better-established Wald \emph{et al.} formulation using their, now standardized, notation. A concrete and rigorous discussions can be found in the original papers, \eg in Ref. \cite{Wald:1999wa}, while for a  pedagogical review we refer to Ref. \cite{Hajian:2015eha}. 
\paragraph{Covariant phase space.} Phase space is a manifold $\mathcal{M}$ which is equipped with a symplectic two-form $\Omega$.  To construct $\Omega$ for a given $d$-dimensional generally invariant theory, one may start through the action
\begin{equation}
\mathcal{S}[\Phi]=\int \mathbf{L},
\end{equation}
where $\mathbf{L}$ is the Lagrangian $d$-form, and $\Phi$ is used to denote collectively all the dynamical fields. Variation of the action leads to
$$\delta \mathbf{L}[\Phi]=\mathbf{E}_{\Phi}\delta\Phi+\mathrm{d}\mathbf{\Theta}_{_\text{LW}}(\delta\Phi,\Phi),$$
where $\delta\Phi$ is a generic field perturbation and provides the basis for tangent space of the phase space ${\cal M}$. The set of equations $\mathbf{E}_{\Phi}=0$ gives the equations of motion. $\mathbf{\Theta}_{_\text{LW}}$ is a $d\!-\!1$-form over the spacetime and one-form on the tangent bundle of ${\cal M}$, \emph{i.e.} $\mathbf{\Theta}_{_\text{LW}}$ is a $(d-1; 1)$-form. On-shell equality is denoted by $\approx$:
\begin{equation}\label{delta L Theta}
\delta \mathbf{L}[\Phi]\approx\mathrm{d}\mathbf{\Theta}_{_\text{LW}}(\delta\Phi,\Phi)\,.
\end{equation}
A  symplectic two-form in the CPSM, the {Lee-Wald} symplectic form \cite{Lee:1990gr}, is defined as
\begin{equation}\label{Omega LW}
\Omega_{_\text{LW}}^{^\Sigma}(\delta_1\Phi,\delta_2\Phi,\Phi)=\int_\Sigma \boldsymbol{\omega}_{_\text{LW}}(\delta_1\Phi,\delta_2\Phi,\Phi)\, 
\end{equation}
where
\begin{equation}\label{omega LW}
\boldsymbol{\omega}_{_\text{LW}}(\delta_1\Phi,\delta_2\Phi,\Phi)\equiv\delta_1\mathbf{\Theta}_{_\text{LW}}(\delta_2\Phi,\Phi)-\delta_2\mathbf{\Theta}_{_\text{LW}}(\delta_1\Phi,\Phi)\,,
\end{equation}
where $\boldsymbol{\omega}_{_\text{LW}}$ is a $(d-1; 2)$-form and $\delta_{1}\Phi$ and $\delta_2\Phi$ are two arbitrary field perturbations which are members of the tangent bundle of the $\mathcal{M}$. In Eq. \eqref{Omega LW}, $\Sigma$ is a  codimension-1 spacelike surface. Hereafter, we restrict to solutions of equations of motion (up to gauge transformations), with the tangent bundle being perturbations satisfying linearized field equations around $\Phi$,  for which \cite{Lee:1990gr, Wald:1999wa}
\begin{enumerate}
\item $\mathrm{d}\boldsymbol{\omega}_{_\text{LW}}(\delta_1\Phi,\delta_2\Phi,\Phi)\approx0$,
\item $\Omega_{_\text{LW}}^{^\Sigma}$ has no degenerate directions and is conserved and independent of $\Sigma$, if  $\Sigma$ is a Cauchy surface with appropriate boundary conditions for fields and their perturbations on $\partial\Sigma$.
\end{enumerate}
Therefore, if we denote the  set of solutions to equations of motion (up to gauge transformations) by ${\cal F}$, $({\cal F}; \Omega)$ constitutes a well-defined phase space. The tangent space of ${\cal F}$, which is spanned by $\delta\Phi$, will be denoted by $T_{\cal F}$.

\paragraph{Freedom/ambiguities on $\boldsymbol\omega$.} The Lee-Wald (pre)symplectic form $\boldsymbol{\omega}_{_\text{LW}}$ is constructed from the Lagrangian $d$-from. Eq.\eqref{delta L Theta} reveals that one has a freedom in defining $\mathbf{\Theta}$ up to  a $(d-2; 1)$-form ``boundary terms''  $\mathbf{Y}$:
\be\label{Theta-Y}
\mathbf{\Theta}_{_\text{LW}}(\delta \Phi,\Phi)\to \mathbf{\Theta}(\delta \Phi,\Phi)=\mathbf{\Theta}_{_\text{LW}}(\delta \Phi,\Phi)+\mrd \mathbf{Y}(\delta\Phi,\Phi).
\ee
As a result, the symplectic current has the following  ambiguity:
\begin{equation}
\boldsymbol{\omega}_{_\text{LW}}(\delta_1\Phi,\delta_2\Phi,\Phi)\to \boldsymbol{\omega}(\delta_1\Phi,\delta_2\Phi,\Phi)=\boldsymbol{\omega}_{_\text{LW}}(\delta_1\Phi,\delta_2\Phi,\Phi)+\mrd \big[\delta_1\mathbf{Y}(\delta_2\Phi,\Phi)-\delta_2\mathbf{Y}(\delta_1\Phi,\Phi)\big]\,.
\end{equation}
Some specific choices for the boundary term $\mathbf{Y}$ have been discussed in the literature; for example,  if we base the formulation on equations of motion and not the action, we get a specific $\mathbf{Y}$ term {and the corresponding ``invariant symplectic form'' $\boldsymbol{\omega}_{_{\text{inv}}}$ \cite{Barnich:2001jy,Barnich:2007bf}}. Other choices of boundary terms may be chosen and used in different contexts, \eg those motivated by the AdS/CFT \cite{Compere:2008us} or possibly by different boundary conditions on fields and their perturbations \cite{Compere:2014cna, Compere:2015knw,CHSS:2015mza,CHSS:2015bca}.

However, being linear in $\delta_\epsilon\Phi$,  for the exact symmetries the boundary term relating  $\boldsymbol{\omega}_{_\text{LW}}$ and $\boldsymbol{\omega}_{_\text{inv}}$ vanishes \cite{Barnich:2007bf, Wald:1999wa} and hence as long as conserved charges associated with the exact symmetries are concerned, these two lead to the same result.

\paragraph{Conserved charges on the phase space.} On may use $\boldsymbol{\omega}$ to define conserved charges associated with a specific set of transformations generated by $\epsilon$, $\delta_\epsilon\Phi$. The transformations we consider here are among local (gauge) transformations, which also include the exact symmetries.  Since $\mathrm{d}\boldsymbol{\omega}(\delta_1\Phi,\delta_2\Phi,\Phi)\approx0$ for any two perturbations satisfying linearized equations of motion, locally and on the phase space ${\cal F}$, 
\begin{equation}\label{LW dk}
\boldsymbol{\omega}(\delta\Phi,\delta_\epsilon\Phi,\Phi)\approx \mathrm{d}\boldsymbol{k}_\epsilon(\delta\Phi, \Phi),
\end{equation}
where $\boldsymbol{k}_\epsilon$ is a $(d-2; 1)$-form. The above is called the fundamental identity of the CPSM.

One may integrate $\boldsymbol{k}_\epsilon$ to define perturbations of charge associated with $\epsilon$ \cite{Iyer:1994ys}:
\begin{equation}\label{delta H via k}
 \delta H_{\epsilon}(\Phi)= \int_\Sigma \mrd \boldsymbol{k}_\epsilon(\delta\Phi,\Phi)=\oint_{\partial \Sigma}\boldsymbol{k}_\epsilon(\delta\Phi,\Phi)\,.
\end{equation} 

\paragraph{Example: Charges associated with diffeomorphisms+gauge transformations.} In the presence of some number of $U(1)$ gauge fields $A^a$, labelled by index $a$, besides diffeomorphisms generated by $\xi$'s, the Lagrangian can also be gauge invariant, \emph{i.e.} $\mathbf{L}\to\mathbf{L}$ under $A^a\to A^a+\mrd\lambda^a$ for arbitrary scalars $\lambda^a$. Our $\delta_\epsilon$  now involves both diffeomorphisms and gauge transformations, $\epsilon\equiv \{\xi,\lambda^a\}$, explicitly, $\delta_\epsilon\Phi=\{\mathscr{L}_\xi\Phi,\delta_{\lambda^a} \Phi\}$,

The charge perturbations (also called Hamiltonian generators) associated with the variations generated by $\epsilon$,  $\delta H_\epsilon(\Phi)$, are then given as
\begin{align}\label{delta H xi}
\delta H_{\epsilon}(\Phi)&\equiv \Omega(\delta\Phi,\delta_\epsilon\Phi,\Phi)=\int_\Sigma \boldsymbol{\omega}(\delta\Phi,\delta_\epsilon\Phi,\Phi)\\
&=\int_\Sigma \Big(\delta\mathbf{\Theta}(\delta_\epsilon\Phi,\Phi)-\delta_\epsilon\mathbf{\Theta}(\delta\Phi,\Phi)\Big)=\oint_{\partial\Sigma}\boldsymbol{k}_{\epsilon}(\delta\Phi,\Phi) \,. \label{omega in terms of theta}
\end{align} 
If the above integrals are finite and nonvanishing,  $\delta H_{\epsilon}(\Phi)$ then corresponds to a conserved charge variation. 
The generic form of $\boldsymbol{k}_\epsilon$  is (see Appendix \ref{appen-A})
\begin{equation}\label{k_xi general}
\boldsymbol{k}_\epsilon(\delta\Phi,\Phi)=\delta \mathbf{Q}_\epsilon-\xi \cdot \mathbf{\Theta}(\delta \Phi,\Phi)\,,
\end{equation} 
in which $\mathbf{Q}_\epsilon$ is the \emph{Noether-Wald charge density},
\begin{equation}\label{Noether Wald}
\mrd \mathbf{Q}_\epsilon\equiv \mathbf{\Theta}(\delta_\epsilon\Phi,\Phi)-\xi \! \cdot \! \mathbf{L}\,.
\end{equation}
One may then evaluate the charge knowing how to compute $\boldsymbol{k}_\epsilon$. 

\paragraph{Integrability of charges.}
The charge $H_\epsilon[\Phi]$ is well defined if $\delta H_{\epsilon}(\Phi)$ is \emph{integrable} over the phase space. This condition holds if $(\delta_1\delta_2-\delta_2\delta_1)H_\epsilon (\Phi)=0$, in which $\Phi$s are any field configuration in the presumed phase space ${\cal F}$, and $\delta_{1,2}\Phi$ are any arbitrary chosen member of its tangent bundle. For field-independent transformations, \ie when $\delta\epsilon=0$, straightforward analysis (see Appendix \ref{appen-A}) shows that this condition is equivalent to \cite{Lee:1990gr}
\be\label{integrability-cond}
\oint_{\partial\Sigma} \xi\cdot\boldsymbol{\omega}(\delta_1\Phi,\delta_2\Phi,\Phi)\approx 0, \qquad \forall \Phi\in{\cal F},\ \delta\Phi\in T_{\cal F}.
\ee

\paragraph{Field-dependent  transformations and their charges.} In the standard CPSM which we reviewed above there is an implicit assumption that the generator of transformations $\epsilon$ are field independent, \emph{i.e.} $\delta \epsilon=0$. For the important example of exact symmetries, which is our main focus in this work, however, this assumption is not always respected and one needs to revisit the steps of CPSM given above for possible modifications. This has been discussed in the Appendix of Ref. \cite{Compere:2015knw}, and we give the results here. The ``field-dependence adjusted'' charge variations are 
\begin{align}
\delta H_{\epsilon}&\equiv \int_\Sigma \big(\delta^{[\Phi]}\mathbf{\Theta}(\delta_\epsilon\Phi,\Phi)-\delta_\epsilon\mathbf{\Theta}(\delta\Phi,\Phi)\big)\,. 
\end{align} 
In this definition, $\delta^{[\Phi]}$ in $\delta^{[\Phi]}\mathbf{\Theta}(\delta_\epsilon\Phi,\Phi)$ acts only on $\Phi$ and not on the $\epsilon$ inside $\mathbf{\Theta}$. Hence, one can still compute $\delta H_{\epsilon}$ via
$\delta H_{\epsilon}= \oint_{\partial \Sigma}\boldsymbol{k}_\epsilon(\delta\Phi,\Phi)$, but the integrability condition is now modified to \cite{Compere:2015knw}
\begin{equation}\label{integrability cond modified}
\oint_{\partial\Sigma} \Big(\xi\cdot \boldsymbol{\omega}(\delta_1\Phi,\delta_2\Phi,\Phi)+\boldsymbol{k}_{\delta_1\epsilon}(\delta_2\Phi,\Phi) -\boldsymbol{k}_{\delta_2\epsilon}(\delta_1\Phi,\Phi)\Big)\approx0.
\end{equation}

{\paragraph{Pure gauge transformations.} If for an $\epsilon$,  $\delta H_\epsilon=0$ over all of the phase space $\mathcal{F}$, then $\epsilon$ is the generator of a ``pure gauge transformation" \cite{Lee:1990gr}. The field configurations related by such coordinate (gauge) transformations are physically equivalent and their Hamiltonian generator is trivial and vanishing over the phase space. As is usual in the context of gauge theories or generally covariant theories, one may then define physical states or observables up to gauge equivalent classes.}

\paragraph{{Symplectic nonexact and exact symmetries.}} By definition the phase space $({\cal F}; \Omega)$ has a nondegenerate symplectic two-form. It may happen that for a subspace of the tangent space (to a given point $\Phi$) the symplectic current $\boldsymbol{\omega}$ vanishes on-shell, \emph{i.e.} for a subclass of $\delta_\epsilon\Phi$'s, which will be denoted by $\delta_{\boldsymbol{\omega}}\Phi$,  
\begin{equation}\label{symplectic-condition}
\boldsymbol{\omega}(\delta\Phi,\delta_{\boldsymbol{\omega}}\Phi,\Phi)\approx 0\,,
\end{equation}
for all $\delta \Phi$ satisfying linearized equations of motion. Two important properties follow from \eqref{symplectic-condition}: {(i) conservation, \emph{i.e} independence of $\Sigma$, would be guaranteed for the $\delta H_{\boldsymbol{\omega}}$;} (ii) Independence of the charge on the codimension-2 integration surface $\partial\Sigma$ \cite{Compere:2014cna, CHSS:2015mza, CHSS:2015bca, Hajian:2015eha, Compere:2015knw}. In this case, instead of the usual notion of asymptotic symmetries, we are dealing with charges which may be defined ``everywhere.'' The transformations generated by $\delta_{\boldsymbol{\omega}}\Phi$ are called \emph{symplectic symmetries} \cite{CHSS:2015mza}. 

Eq.\eqref{symplectic-condition} may be satisfied for two classes of field perturbations, the nonexact symmetries $\delta_\chi\Phi$ and the exact symmetries $\delta_\eta\Phi$:
\begin{enumerate}
\item \textbf{Symplectic nonexact symmetries} are the subset of $\delta_{\boldsymbol{\omega}}\Phi$, which are nonzero at least on one point of the phase space. We will denote symplectic nonexact symmetry perturbation by $\delta_\chi\Phi\neq 0.$ Based on a given solution $\Phi_0$, one may construct a full phase space by successive action of $\delta_\chi$ on $\Phi_0$ (which may formally be denoted as $e^{\int_\gamma\delta_\chi}\ \Phi_0$, where $\gamma$ denotes an arbitrary path in the phase space). The elements in this phase space which may be denoted by ${\cal F}_{\Phi_0}$ are then labelled by conserved charges associated with symplectic nonexact symmetries (which are defined by integration of symplectic current over a \emph{generic} smooth, closed, and compact codimension-2 surface $\partial\Sigma$); they fall into the ``orbits'' of the algebra of symplectic charges. For cases when $\Phi_0$ is a locally AdS$_3$ geometry or a near-horizon extremal geometry, respectively see Refs. \cite{Compere:2015knw,CHSS:2015mza,CHSS:2015bca} for discussion on sympelctic charges and their algebra. 
\item \textbf{Symplectic exact symmetries} are the subset of $\delta_{\boldsymbol{\omega}}\Phi$, which vanish all over the phase space. We will denote symplectic exact symmetry perturbation by $\delta_\eta\Phi=0$, for which obviously \eqref{symplectic-condition} is satisfied.
\end{enumerate}
The symplectic exact symmetries, which we will call them exact symmetries hereafter,  are the class of perturbations we will analyze in more detail in the rest of this work. We construct the charges perturbations associated with the exact symmetries, caused through parametric variations $\hat\delta\Phi$, by focusing on $\boldsymbol{k}_\eta(\hat\delta\Phi,\Phi)$. We will show below that these perturbations form their own phase space which will be denoted as ${\cal F}_p$.

\section{Solution phase space and associated conserved charges}\label{sec-3}

In the previous section we introduced the machinery of the covariant phase space method through which one can in principle compute conserved charges variations. In our notation we used $\delta\Phi$ for any field perturbation which satisfies linearized equations of motion. In the class of theories with local gauge symmetries (like diffeomorphism invariant theories or the Maxwell theory), a sector of such field perturbations is $\delta_\epsilon\Phi$'s which denote field perturbations under generic gauge transformation, generated by $\epsilon$. It may happen that one can associate well-defined (finite and nontrivial) charge variations to a subclass of  $\delta_\epsilon\Phi$'s (which, \eg are allowed by a specific boundary condition or keep a given gauge intact). There is yet another subclass of the gauge transformations, generated by $\eta$, for which $\delta_\eta\Phi=0$. The latter is called exact symmetries. 

Besides $\delta_\epsilon\Phi$, we may consider a different category of perturbations/variations, the parametric variations $\hat\delta\Phi$. In this section we will focus on this class of variations and the class of exact symmetries $\delta_\eta\Phi$.

\subsection{Parametric variations and solution phase space}

Let us consider family of solutions collectively denoted by $\Phi$. Each solution may be identified with a set of parameters $p_\alpha$: $\Phi=\Phi(x; p_\alpha).$ {As is usual, any such solution is defined up to some coordinate or gauge transformation.} As the notation  indicates, here we are assuming that $p_\alpha$'s are continuous real labels on the solution field configuration $\Phi$. In our analysis here, we do not consider discrete labels and also assume that for each given set of $p_\alpha$'s we have a distinct solution. Explicitly and by definition, $\delta_\epsilon\Phi$ will not take a solution with given $p_\alpha$'s to another solution with different $p_\alpha$'s. 

The parametric variations $\hat\delta\Phi$ \cite{HSS:2014twa} are then defined as
\begin{equation}\label{parametric variations}
\hat{\delta}\Phi\equiv \frac{\partial {\Phi}}{\partial p_\alpha}\delta p_\alpha.
\end{equation}  
{Note that \eqref{parametric variations} defines $\hat\delta\Phi$ up to pure gauge transformations. }
It is clear from the definition that they satisfy linearized equations of motion because $\Phi=\Phi(x; p_\alpha)$ and $\Phi+\hat\delta\Phi=\Phi(x; p_\alpha+\delta p_\alpha)$ are both solutions to  $\mathbf{E}_{{\Phi}}=0$ and hence   $\frac{\delta\mathbf{E}(\Phi)}{\delta\Phi}\big|_{\Phi} \hat\delta\Phi=0$. With the above discussion, it becomes clear that the set of parametric variations $\hat{\delta}\Phi$ and those associated with { nontrivial (nonpure)} gauge transformations $\delta_\epsilon\Phi$  only overlap at $\delta\Phi= 0$ cases, \emph{i.e.} on the set of exact symmetries. We shall use this fact in our construction below.

From the discussions of previous sections, we learn that tangent space of the covariant phase space manifold $\mathcal{F}$ has different subspaces, \eg those spanned by $\delta_\epsilon\Phi$ or those with $\hat\delta\Phi$.  The set of solutions ${\Phi}(x^\mu; p_\alpha)$ with tangent space restricted to $\hat\delta\Phi$ constitutes a (sub)manifold of ${\cal F}$, which can be denoted by ${\mathcal{F}_p}$. One may then show that the symplectic two-form  $\hat\Omega$ defined as
\begin{equation}\label{Omega LW hat}
\hat{\Omega}\equiv\Omega(\hat\delta_1\Phi,\hat\delta_2\Phi,\Phi)=\int_\Sigma \boldsymbol{\omega}(\hat\delta_1\Phi,\hat\delta_2\Phi,\Phi)\, 
\end{equation}
where
\begin{equation}\label{omega LW hat}
\boldsymbol{\omega}(\hat\delta_1\Phi,\hat\delta_2\Phi,\Phi)=\hat\delta_1\mathbf{\Theta}(\hat\delta_2\Phi,\Phi)-\hat\delta_2\mathbf{\Theta}(\hat\delta_1\Phi,\Phi)\,,
\end{equation}
constitute a well-defined symplectic two-form on ${\cal F}_p$; \emph{i.e.} $({\cal F}_p; \hat\Omega)$ forms a phase space, which we will dub as the \emph{solution phase space}. 
Note that the symplectic structure above already includes the boundary $\mathbf{Y}$ terms  discussed in the previous section.

\subsection{Conserved charges on solution phase space} 
One may readily use the general covariant phase space method formulation of computing conserved charges, to the solution phase space. To this end, we first define charge variations and then integrating them over a given path in ${\cal F}_p$, we define the charges.

\paragraph{Conserved charge variations:} Charge variations associated with perturbations $\epsilon$ can be calculated by inserting $\hat{\delta}\Phi$ into the right-hand side of \eqref{delta H via k},
\begin{equation}\label{hat delta H}
 \hat\delta H_{\epsilon}= \oint_{\partial \Sigma}\boldsymbol{k}_\epsilon(\hat\delta\Phi,\Phi)\,.
\end{equation}
Hence, by this equation one is able to calculate $\hat\delta H_\epsilon$ on a solution $\Phi$ for any given covariant theory identified by $\mathbf{L}$. Thanks to the linearity of $\hat\delta H_{\epsilon}$ in $\hat\delta\Phi$, one can insert \eqref{parametric variations} into \eqref{hat delta H} term by term, so the explicit computations can be performed easier. 
We note that in \eqref{hat delta H}, in general $\hat \delta \epsilon\neq0$; \emph{i.e.} we may have \emph{parameter-dependent} transformations.

\paragraph{Integrability:} In order for the charge variations $\hat \delta H_\epsilon$ to be well-defined over ${\cal F}_p$, we need to require 
\begin{equation}\label{integrability cond base parametric}
(\hat\delta_1\hat\delta_2-\hat\delta_2\hat\delta_1)H_\epsilon=0\,.
\end{equation}
The above then leads to the integrability condition for parametric variations, 
 \begin{equation}\label{integrability cond para}
\oint_{\partial\Sigma} \Big(\xi\cdot \boldsymbol{\omega}(\hat\delta_1\Phi,\hat\delta_2\Phi,\Phi)+\boldsymbol{k}_{\hat\delta_1\epsilon}(\hat\delta_2\Phi,\Phi) -\boldsymbol{k}_{\hat\delta_2\epsilon}(\hat\delta_1\Phi,\Phi)\Big)\approx0\,.
\end{equation}
The above integrability condition will be used to constrain transformation generators  $\epsilon$. We note that \eqref{integrability cond para} is linear and hence  if $\epsilon_1$ and $\epsilon_2$ are two generators with integrable conserved charges, then $c_1\epsilon_1+c_2\epsilon_2$ (for any $c_1,c_2$ which are constant on spacetime and on the solution space), would be also integrable. 
{\paragraph{Pure gauge transformations among $\hat\delta\Phi$.} If for an $\epsilon$, the $\hat \delta H_\epsilon=0$ over all of the solution phase space $\mathcal{F}_p$, then we consider the $\epsilon$ as generator of a ``pure gauge transformation." Recalling the presence of such pure gauge transformations, implies that the solution phase is an infinite dimensional phase space.}

\paragraph{Integrating on solution phase space:} The calculated $\hat\delta H_{\epsilon}$ in \eqref{hat delta H} would be a function of parameters, $\hat\delta H_{\epsilon}(p_\alpha)$. It might or might not be a well-defined and integrable function. In the case of well-defined and integrable $\hat\delta H_{\epsilon}$, it can be integrated over the solution phase space  which is parametrized by $p_\alpha$'s. Hence, by a choice of the appropriate reference point, $H_{\epsilon}(p_\alpha)$ can be found. More precisely
\begin{equation}\label{finite H xi}
H_\epsilon[\Phi]=\int_{\bar p}^{p} \hat\delta H_\epsilon+H_\epsilon[\bar\Phi]\,,
\end{equation} 
in which the integration is performed over arbitrary integral curves which connect a reference field configuration $\bar\Phi$ to the $\Phi$ on the solution phase space {(studied but not been well appreciated in Refs.  \cite{Barnich:2003xg} and\cite{Wald:1999wa})}. The $H_\epsilon[\bar\Phi]$ is the reference point for the $H_\epsilon$ defined on the reference field configuration $\bar\Phi$.

\paragraph{Specializing to exact symmetries.} Although our formulation above may be used for any gauge or diffeomorphism $\delta_\epsilon\Phi$, in what follows we will be using it only for exact symmetries generated by $\eta$. In our notations the exact symmetries $\eta$ include a Killing vector (isometry) of the geometry, which will be generically denoted by $\zeta$, and/or an internal gauge transformation generated by $\lambda$: $\eta=\{\zeta,\lambda\}$. It may happen that for specific Killing vectors $\zeta$, $\lambda$ is also fixed in terms of $\zeta$.

As  pointed out, exact symmetries are the only place where the parametric variations $\hat{\delta}\Phi$ and the gauge symmetry ones $\delta_\epsilon\Phi$ overlap. In this sense, the charge variations $\hat\delta H_\eta[\Phi]$ may be defined  on the solution phase space of which the tangent space is only covered by $\hat{\delta}\Phi$. To compute the charges for exact symmetries, we first compute $\boldsymbol{k}_\eta(\hat\delta\Phi,\Phi)$ for a given set of solutions with parameters $p_\alpha$ and parametric variations $\hat\delta\Phi$. $\boldsymbol{k}_\eta$ is a $d-2$-form on spacetime and a one-form over the solution phase space, and one may integrate over smooth, codimension-2, compact spacelike surfaces $\partial\Sigma$ to obtain charge variations $ \hat\delta H_{\eta}$ which are one-forms over the solution phase space. If the charge is integrable, one may then define the charge $H_\eta[\Phi]$ by integrating the charge variation over an (appropriately chosen) path in the solution phase space. Integrability then implies that the charge should be path independent.

Finally we point out that dealing with the exact symmetries for which $\delta_\eta\Phi=0$, brings the interesting result that the boundary $\mathbf{Y}$ term (which is an ambiguity in the definition of $\boldsymbol{\omega}$) would not contribute to the charges, and hence for the computation of conserved Noether charges, one may only focus on the Lee-Wald (LW) (pre)sympletic structure density $\boldsymbol{\omega}_{_\text{LW}}$. Therefore, in what follows, to simplify the notation, we will only consider the Lee-Wald symplectic structure density and simplify the notation by dropping the LW index.

\section{Example: conserved charges  in Einstein-Maxwell-Scalar-$\mathbf{\Lambda}$ theory}\label{sec-4}

As reviewed and explained above, $\boldsymbol{k}_\eta$ depends on the details of the theory. As a concrete example and to illustrate how our method works,  let us consider  Einstein-Maxwell-scalar (EMS) theories, with cosmological constant $\Lambda$. The dynamical fields $\Phi$ would be the metric $g_{\mu\nu}$, some number of Abelian one-form gauge  fields $A^a$ and some number of scalar fields $\phi^I$ governed by the Lagrangian
\begin{equation}\label{Lagrangian scalar}
\mathcal{L}=\frac{1}{16\pi G}\big(R-2\mathrm{f}_{_{IJ}}\nabla^\mu\phi^I\nabla_\mu\phi^J-\mathrm{k}_{ab}F_{\mu\nu}^a F^{b\, \mu\nu}-V(\phi)\big)\,.
\end{equation}
$R$ is the Ricci scalar, and $F^a=\mrd A^a$ is the field strength. $\mathrm{f}_{_{IJ}}$ and $\mathrm{k}_{ab}$ are some functions of $\phi$. The Lagrangian $d$-form is the Hodge dual of \eqref{Lagrangian scalar}, $\mathbf{L}=\star \mathcal{L}$,
\begin{equation}\label{Lagrangian top form}
\mathbf{L}=\frac{\sqrt{-g}}{d!}\,\,\epsilon_{\mu_1\mu_2\cdots \mu_d}  \,\mathcal{L}\,\,\mrd x^{\mu_1}\wedge \mrd x^{\mu_2}\wedge\cdots\wedge \mrd x^{\mu_d}\,.
\end{equation}
where $\epsilon_{\mu_1\mu_2\cdots \mu_d}$ is the Levi-Civita symbol, \emph{i.e.} $\epsilon_{_{012\cdots d-1}}=+1$ and changes sign according to the permutations of indices. The surface $d\!-\!1$-form  $\mathbf{\Theta}$ would be 
\begin{equation}
\mathbf{\Theta}=\frac{\sqrt{-g}}{(d-1)!}\,\,\epsilon_{\mu\mu_1\cdots \mu_{d-1}} \,(\Theta^{\text{E}\,\mu}+\Theta^{\text{M}\,\mu}+\Theta^{\text{S}\,\mu})\,\,\mathrm{d}x^{\mu_1}\wedge \cdots\wedge \mathrm{d}x^{\mu_{d-1}} 
\end{equation}
in which
\begin{align}
&\Theta^{\text{E}\,\mu}(\delta\Phi,\Phi)=\frac{1}{16\pi G}(\nabla_\nu h^{\mu\nu}-\nabla^\mu h)\,, \label{EH Theta}\\
&\Theta^{\text{M}\,\mu}(\delta\Phi,\Phi)=\frac{-1}{4\pi G}\mathrm{k}_{ab}\,F^{a\,\mu\nu}\,\delta A^b_{\nu}\,,\\
&\Theta^{\text{S}\,\mu}(\delta\Phi,\Phi)=\frac{-1}{4\pi G}\mathrm{f}_{_{IJ}}\,\nabla^\mu\phi^I\delta\phi^J\,.
\end{align}
where $h^{\mu\nu}\equiv g^{\mu \sigma}g^{\nu\tau}\delta g_{\sigma \tau}$ and $h\equiv h^\mu_{\,\,\mu}$. For  $\epsilon=\{\xi,\lambda^a\}$, the Noether-Wald $d\!-\!2$-form $\mathbf{Q}_\epsilon$ can be read through \eqref{Noether Wald} as 
\begin{equation}
\mathbf{Q}_\epsilon=\frac{\sqrt{-g}}{(d-2)!\,2!}\,\,\epsilon_{\mu\nu\mu_1\cdots \mu_{d-2}} \,(\mathrm{Q}_\epsilon^{{\text{E}\,\,\mu\nu}}+\mathrm{Q}_\epsilon^{{\text{M}\,\,\mu\nu}}+\mathrm{Q}_\epsilon^{{\text{S}\,\,\mu\nu}})\,\,\mathrm{d}x^{\mu_1}\wedge \cdots\wedge \mathrm{d}x^{\mu_{d-2}} 
\end{equation}
 in which
\begin{align}
& \mathrm{Q}_\epsilon^{{\text{E}\,\,\mu\nu}}=\frac{-1}{16\pi G}(\nabla^\mu\xi^\nu-\nabla^\nu\xi^\mu)\,,\\
& \mathrm{Q}_\epsilon^{{\text{M}\,\,\mu\nu}}=\frac{-1}{4\pi G}\mathrm{k}_{ab}F^{a\,\,\mu\nu}(A^b_\rho\xi^\rho+\lambda^b)\,,\\
&\mathrm{Q}_\epsilon^{{\text{S}\,\,\mu\nu}}=0\,.
\end{align}
Next, we should compute $d\!-\!2$-form $\boldsymbol{k}_\epsilon(\delta\Phi,\Phi)$, which can be done using  \eqref{k_xi general}. The calculations are cumbersome but straightforward \cite{Geoffrey:etal} (\emph{e.g.} see Appendix C.6 in Ref. \cite{Hajian:2015eha} for the Einstein-Hilbert theory). The final result is 
\begin{equation}
\boldsymbol{k}_\epsilon(\delta\Phi,\Phi)=\frac{\sqrt{-g}}{(d-2)!\,2!}\,\,\epsilon_{\mu\nu\mu_1\cdots \mu_{d-2}} \,(k_\epsilon^{\text{E}\,\mu\nu}+k_\epsilon^{\text{M}\,\mu\nu}+k_\epsilon^{\text{S}\,\mu\nu})\,\,\mathrm{d}x^{\mu_1}\wedge \cdots\wedge \mathrm{d}x^{\mu_{d-2}} 
\end{equation}
where
\begin{align}
&k_\epsilon^{\text{E}\,\mu\nu}(\delta\Phi,\Phi)=\dfrac{1}{16 \pi G}\Big(\Big[\xi^\nu\nabla^\mu h
-\xi^\nu\nabla_\sigma h^{\mu\sigma}+\xi_\sigma\nabla^{\nu}h^{\mu\sigma}+\frac{1}{2}h\nabla^{\nu} \xi^{\mu}-h^{\rho\nu}\nabla_\rho\xi^{\mu}\Big]-[\mu\leftrightarrow\nu]\Big)\,,\label{EH k}\\
&k_\epsilon^{\text{M}\,\mu\nu}(\delta\Phi,\Phi)=\frac{1}{8 \pi G}\Big(\Big[\big(\frac{-h}{2} \,\mathrm{k}_{ab}\, F^{a\,\mu\nu}\!+\!2\,\mathrm{k}_{ab}\,F^{a\,\mu\rho}h_\rho^{\;\;\nu}-\mathrm{k}_{ab}\,\delta F^{a\,\mu\nu}\!-\!\frac{\partial\,\mathrm{k}_{ab}}{\partial \phi^I}\,F^{a\,\mu\nu}\delta\phi^I\big)({\xi}^\rho A^b_\rho+\lambda^b)\hspace*{3cm}\nonumber\\
 &\hspace*{4cm} - \,\mathrm{k}_{ab}\,F^{a\,\mu\nu}\xi^\rho \delta A^b_\rho-2\,\mathrm{k}_{ab}\,F^{a\,\rho\mu}\xi^\nu \delta A^b_\rho\Big]-[\mu\leftrightarrow\nu]\Big)\,,\label{k for EM} \\
 &k_\epsilon^{\text{S}\,\mu\nu}(\delta\Phi,\Phi)=\frac{1}{8\pi G}\Big(\xi^\nu\,\mathrm{f}_{_{IJ}}\,\nabla^\mu\phi^I\,\delta\phi^J-[\mu\leftrightarrow\nu]\Big)\,.
\end{align} 
for any chosen cosmological constant $\Lambda$.

So far $\delta\Phi$ and $\epsilon$ were generic; hereafter we restrict $\delta\Phi$ to parametric variations $\hat\delta\Phi$ and $\epsilon$ to exact symmetries $\eta$. To illustrate how our formulation works, we analyze some examples. They are chosen as simple as possible, but rich enough to be used for our purpose. Some further examples may be found in Ref. 
\cite{Hajian:2015eha}.

\subsection{Charges associated with Kerr-AdS black hole}\label{sec-4-1}

As our first example we start with the Kerr-AdS black hole, which is an aymptotically AdS solution to the Einstein-Hilbert theory, with negative cosmological constant $\Lambda$. The Lagrangian density for this theory is $\mathcal{L}\!=\!\frac{1}{16\pi G}(R-2\Lambda)$,  with the metric as its only dynamical field. The metric of the Kerr-AdS black hole in coordinates which is nonrotating at infinity, is given as \cite{Carter:1968ks}
\begin{align}\label{Kerr-AdS metric}
\hspace{-0.23cm}\mathrm{d}s^2= -&\Delta_\theta(\frac{1+\frac{r^2}{l^2}}{\Xi}\!-\!\Delta_\theta f)\mathrm{d}t^2+\frac{\rho ^2}{\Delta_r}\mathrm{d}r^2+\frac{\rho ^2}{\Delta_\theta} \mathrm{d}\theta ^2 -2\Delta_\theta fa\sin ^2 \theta\,\mathrm{d}t \mathrm{d}\varphi\nonumber \\
+&\left( \frac{r^2+a^2}{\Xi}+fa^2\sin ^2\theta \right)\sin ^2\theta\,\mathrm{d}\varphi ^2\,,
\end{align}
where
\begin{align}\label{Kerr-AdS entities}
\rho^2 &\equiv r^2+a^2 \cos^2 \theta\,,\qquad \Delta_r \equiv (r^2+a^2)(1+\frac{r^2}{l^2})-2Gmr\,,\nonumber\\
\Delta_\theta&\equiv 1-\frac{a^2}{l^2}\cos ^2\theta\,,\qquad
f\equiv\frac{2Gmr}{\rho ^2\Xi^2}\,,\qquad \Xi\equiv 1-\frac{a^2}{l^2}\,.
\end{align}
The solution is specified by  two  parameters  $m$ and $a$. The radius of the AdS$_4$ has been denoted by $l$, which is related to the $\Lambda$ by $\Lambda=\dfrac{-3}{l^2}$. The charges and Smarr relation for this solution have been in particular studied and analyzed in Refs. \cite{Gibbons:2004ai, Barnich:2004uw,Olea:2005gb}.

To employ our formulation, we start with parametric variations for this solution, which  explicitly are
\begin{equation}\label{Kerr AdS parametric var}
\hat\delta g_{\mu\nu}=\frac{\partial g_{\mu\nu}}{\partial m}\delta m+\frac{\partial g_{\mu\nu}}{\partial a}\delta a\,.
\end{equation}
In the absence of gauge fields, the exact symmetries are identical to Killing vectors $\zeta$. The metric has two obvious Killings $\partial_t, \partial_\varphi$ and of course any linear combination of these two with arbitrary $m,a$-dependent coefficients is also a Killing. Nonetheless, as we will see, not any combination of these Killings leads to integrable charges. 

\paragraph{Mass:} One can calculate $\hat{\delta}H_{\partial_t}$ using $\boldsymbol{k}_\xi^{\mathrm{E}}$ from \eqref{EH k}, and choosing $\partial\Sigma$ to be \emph{any} $(d\!-\!2)$-dimensional, spacelike, smooth and closed surface which surrounds the black hole (includes $r=0$). For simplicity, one can choose $\partial \Sigma$ to be constant $t, r$ surfaces, with arbitrary $t,r$. {Using \eqref{Kerr AdS parametric var} as perturbations in \eqref{EH k}, the result would be }
\begin{align}\label{Kerr AdS m}
\hat{\delta}H_{\partial_t}=\frac{1}{\Xi^2}\delta m+\frac{4ma}{\Xi^3}\delta a \,.
\end{align}
One may then check that $\hat{\delta}H_{\partial_t}$ is integrable over the solution phase space, by a direct check of \eqref{integrability cond para}. Explicitly, the Killing $N\partial_t$ with normalization $N=1$ leads to an integrable charge. Alternatively, one may observe that $\hat{\delta}H_{\partial_t}$ \eqref{Kerr AdS m} is a variation of a function on the parameters $m,a$. That function is explicitly the $\frac{m}{\Xi^2}+{\rm const.}$, where the const. should be a constant over the solution phase space; \emph{i.e.}  it should be independent of the parameters of the solution. However, it could be a function of the parameters of the theory, like $\ell$ or $G$. We fix this constant by choosing the pure AdS$_4$ spacetime ($m=a=0$ case) as the reference point with $H_{\partial_t}[\bar\Phi]=0$, and we find the ``mass'' of the Kerr-AdS black hole, 
\begin{equation}
M\equiv H_{\partial_t}=\frac{m}{\Xi^2}\,.
\end{equation}
\paragraph{Angular momentum:} Similar calculation for the Killing $\partial_{\varphi}$, on any surface of constant $t,r$ surrounding the hole, (considering the standard additional minus sign) leads to 
\begin{equation}
-\hat{\delta}H_{\partial_\varphi}=\frac{a}{\Xi^2}\delta m+\frac{m(1+3\frac{a^2}{l^2})}{\Xi^3}\delta a\,,
\end{equation}
which is also integrable over the parameters.\footnote{This means that the charge of $N\partial_\varphi$ with normalization $N=1$ is integrable over the solution phase space.} Reference point $H_{\partial_\varphi}[\bar\Phi]$ can be chosen to vanish for pure AdS, i.e., on the solution with $m=a=0$. Hence, the  angular momentum for the Kerr-AdS black hole is found to be
\begin{equation}
J\equiv -H_{\partial_\varphi}=\frac{ma}{\Xi^2}\,.
\end{equation}
\paragraph{Entropy:} The Killing vector generating the (Killing) horizon of the black hole  is 
\begin{equation}\label{Kerr-AdS tilde zeta}
\tilde\zeta_{_\mathrm H}=\partial_t+\Omega_{_\mathrm H}\partial_\varphi, \qquad  
\Omega_{_\mathrm H}=\frac{a(1+\frac{r_{_\mathrm H}^2}{l^2})}{r_{_\mathrm H}^2+a^2}
\end{equation}
and $r_{_\mathrm H}$ is the radius of its event horizon and a solution to $\Delta_r=0$, explicitly
\be\label{r-H-Kerr-AdS}
(r_{_\mathrm{H}}^2+a^2)(r_{_\mathrm{H}}^2+{l^2})-2Gm\ell^2 r_{_\mathrm{H}}=0.
\ee
A direct examination of \eqref{integrability cond para} reveals that the charge associated with $\tilde\zeta_{_\mathrm H}$ is \emph{not} integrable. (Note that $\Omega_{_\mathrm H}$ is parameter dependent and hence $\hat{\delta}\tilde \zeta_{_\mathrm H}\neq 0$.) Nonetheless,  one may check that the appropriately normalized vector,  $\zeta_{_\mathrm{H}}\equiv \frac{2\pi}{\kappa}\tilde\zeta_{_\mathrm{H}}$  where $\kappa$ is surface gravity on the horizon,
\begin{equation}
\kappa=\frac{r_{_\mathrm H}(1+\frac{a^2}{l^2}+3\frac{r_{_\mathrm H}^2}{l^2}-\frac{a^2}{r_{_\mathrm H}^2})}{2(r_{_\mathrm H}^2+a^2)}\,,
\end{equation}
is integrable (see Appendix \ref{appen-B-1} for the detailed analysis). Therefore, one may integrate $\hat \delta H_{\zeta_{_\mathrm{H}}}$ to obtain the corresponding charge. We emphasize that $\hat \delta H_{\zeta_{_\mathrm{H}}}$ may be defined over any codimension-2, compact spacelike surface $\partial \Sigma$ which surrounds the hole (not necessarily the horizon). The result which is independent of the choice of such $\partial\Sigma$ is
\begin{equation}\label{Kerr AdS del S}
\hat \delta H_{\zeta_{_\mathrm{H}}}=\frac{\partial \left(\frac{\pi(r_{_\mathrm{H}}^2+a^2)}{G\,\Xi}\right)}{\partial m}\delta m + \frac{\partial \left(\frac{\pi(r_{_\mathrm{H}}^2+a^2)}{G\,\Xi}\right)}{\partial a}\delta a\,.
\end{equation}
{By choosing the pure AdS$_4$ spacetime as the reference point with $H_{\zeta_{_\mathrm{H}}}[\bar\Phi]=0$,} \eqref{Kerr AdS del S} then results in the entropy for the Kerr-AdS black hole \cite{Wald:1993nt,Iyer:1994ys},
\begin{equation}
S\equiv H_{\zeta_{_\mathrm{H}}}=\frac{\pi(r_{_\mathrm{H}}^2+a^2)}{G\,\Xi}\,.
\end{equation}

Hamiltonian generators which are calculated above, are in agreement with the known mass, angular momentum and entropy of the Kerr-AdS black hole \cite{Barnich:2004uw, Gibbons:2004ai}. Before moving to the next example, some comments are in order:
\begin{enumerate}
\item Being on an asymptotic AdS backgorund, the charges defined through other methods (\emph{e.g.} see Ref. \cite{Hollands:2005wt} for a review) may be infinite and need regularization. By contrast in our method, not only are the charges obtained to be finite, but also by an appropriate choice of the reference point, the conserved charges are completely fixed. 
\item Comparing the analysis for $\tilde\zeta_{_\mathrm{H}}$ and $\zeta_{_\mathrm{H}}$ demonstrates the sensitivity of integrability to the normalization of Killing vectors by some function of parameters. More generally, integrability is highly sensitive to the appropriate choice of vector fields. For example, if one had chosen the metric in Boyer-Lindquist coordinates (which can be found \emph{e.g.} in Ref. \cite{Gibbons:2004ai}), the time would be something different than $t$ in \eqref{Kerr-AdS metric}; let us denote it by $\tau$. One may then check that $\partial_\tau$ does \emph{not} have an integrable conserved charge. As a result, the formulation can be used for investigating appropriate Killing vector fields associated with mass or other conserved charges.
\item Entropy has been found as a Hamiltonian generator, instead of the Noether-Wald charge of $\zeta_{_\mathrm{H}}$. Notice that they are different by the last term in \eqref{k_xi general}. That extra term vanishes on the bifurcation of the horizon, because $\zeta_{_\mathrm{H}}$ vanishes there. This is a key point in allowing us to define entropy on \emph{any} surface $\partial\Sigma$ and free us from defining it on the bifurcation surface only, as is prescribed in Wald's entropy \cite{Wald:1993nt}.  Our definition of entropy as \emph{Hamiltonian generator} associated to $\zeta_{_\mathrm{H}}$ is in a sense a more fundamental definition, as we will discuss in Sec. \ref{sec-5}.
\item For the extremal case, where $\kappa=0$, there is no properly normalized Killing vector field $\zeta_{_\text{H}}$. For this case, however, there are infinitely many such Killing vectors in the \emph{near horizon region} \cite{HSS:2013lna}. We will consider one closely related such example in the next subsection.
\end{enumerate}

\subsection{Extremal Kerr-Newman near horizon geometry}\label{sec-4-2} 
For the next example we choose  the near horizon extremal Kerr-Newman geometry. This example is chosen to show how our formulation naturally takes into account the effects of gauge fields, and also for cases which are not a black hole. This geometry is a solution to the Einstein-Maxwell theory with the Lagrangian $\mathcal{L}\!=\!\frac{1}{16\pi G}(R-F^2)$ and is specified through
\begin{align}\label{NHEG metric}
	{\mrd s}^2&=\Gamma\Big(-r^2\mrd t^2+\frac{\mrd r^2}{r^2}+ \mrd\theta^2+\gamma(\mrd\varphi+kr\,\mrd t)^2\Big)\,,\\
\label{NHEG gauge}
A&= f(\mrd\varphi+kr\,\mrd t)-er\,\mrd t\,,
\end{align}
where
\begin{align}\label{NHE KN entities}
\Gamma&=q^2+a^2(1+\cos^2\theta)\,,\qquad \gamma=\left(\frac{q^2+2a^2\sin\theta}{q^2+a^2(1+\cos^2\theta)}\right)^2\,,\qquad k=\frac{2a\sqrt{q^2+a^2}}{q^2+2a^2}\,,\nonumber\\
f&=\frac{-\sqrt{q^2+a^2}\,qa\sin^2\theta}{q^2+a^2(1+\cos^2\theta)}\,,\qquad e=\frac{q^3}{q^2+2a^2}\,.
\end{align}
This solution has two free parameters, $a$ and $q$. It has $SL(2,\mathbb{R})\times U(1)_\varphi$ isometry with the Killing vectors \begin{align}
\xi_- &\!=\!\partial_t\,,\qquad \, \xi_0\!=\!t\partial_t\!-\!r\partial_r,\qquad\,	\xi_+\! =\!\dfrac{1}{2}(t^2\!+\!\frac{1}{r^2})\partial_t\!-\!tr\partial_r-\frac{{k}}{r}{\partial}_{\varphi}\,, \qquad\, \partial_{\varphi}.
\end{align}
Since in our solution, besides the metric we also have a gauge field, not all the above isometries are directly  exact symmetries (of the full solution). 

One may readily see that $\xi_-,\xi_0$ and $\partial_\varphi$ are exact symmetries while  $\xi_+$ is not: under transformations generated by the $\xi_+$, the  solution goes to itself up to an internal $U(1)$ transformation. Explicitly $\delta_{\xi_+}A=\frac{e}{r^2}\mrd r\neq 0$. Nonetheless, one may check that 
\be 
\eta_+=\{\xi_+,\frac{e}{r}+\text{const.}\}
\ee 
is an exact symmetry. The ``const." is a constant on the spacetime, while it can be a function of parameters $q$ and $a$. Its choice should respect the integrability of $\delta H_{\eta_+}$. A consistent choice, as will become apparent below, is to set it to zero.

Let us calculate the conserved charges associated to the mentioned exact symmetries and also the electric charge $Q$ for this geometry. For the Einstein-Maxwell theory, the  $\boldsymbol{k}_\epsilon$ for generic $\epsilon=\{\xi,\lambda\}$ is
\begin{align}\label{NHEKN-k}
&k_\epsilon^{\text{EM}\,\mu\nu}(\delta\Phi,\Phi)=\dfrac{1}{16 \pi G}\Big(\Big[\xi^\nu\nabla^\mu h
-\xi^\nu\nabla_\sigma h^{\mu\sigma}+\xi_\sigma\nabla^{\nu}h^{\mu\sigma}+\frac{1}{2}h\nabla^{\nu} \xi^{\mu}-h^{\rho\nu}\nabla_\rho\xi^{\mu}\Big]-[\mu\leftrightarrow\nu]\Big)\nonumber\\
&+\frac{1}{8 \pi G}\Big(\Big[\big(\frac{-h}{2} F^{\mu\nu}+2F^{\mu\rho}h_\rho^{\;\;\nu}-\delta F^{\mu\nu}\big)({\xi}^\sigma A_\sigma+\lambda)- F^{\mu\nu}\xi^\rho \delta A_\rho-2F^{\rho\mu}\xi^\nu \delta A_\rho\Big]-[\mu\leftrightarrow\nu]\Big)\,.
\end{align} 
{We note in addition that parametric variations for this solution are
\begin{equation}\label{NHEKN-parametric var}
\hat\delta g_{\mu\nu}=\frac{\partial g_{\mu\nu}}{\partial a}\delta a+\frac{\partial g_{\mu\nu}}{\partial q}\delta q\,, \qquad \hat\delta A_{\mu}=\frac{\partial A_{\mu}}{\partial a}\delta a+\frac{\partial A_{\mu}}{\partial q}\delta q\,.
\end{equation}}
\paragraph{Angular momentum.} Choosing $\eta=\{\partial_\varphi,0\}$ as generator of angular momentum, one can insert the parametric variations into the equation above. Then by the equation \eqref{hat delta H}, and choosing $\partial \Sigma$ to be surfaces of constant $t,r$ (only for simplicity of calculations), the result would be
\begin{equation}
-\hat{\delta}H_{\partial_\varphi}=\frac{q^2+2a^2}{G \sqrt{q^2+a^2}}\delta a+\frac{qa}{G\sqrt{q^2+a^2}}\delta q\,.
\end{equation}
As is explicitly seen $\hat{\delta}H_{\partial_\varphi}$ is integrable over the parameters/solution phase space. It is natural to choose  $H_{\partial_\varphi}[\bar\Phi]=0$ for $a=0$. Hence, angular momentum for this geometry is found to be
\begin{equation}
J=\frac{\sqrt{a^2+q^2}}{G}a\,.
\end{equation}
This result is in agreement with the known angular momentum for the extremal Kerr-Newman black hole.
\paragraph{$SL(2,\mathbb{R})$ charges.} For $\eta_-=\{\xi_-,0\}$,  $\eta_0=\{\xi_0,0\}$, and $\eta_+=\{\xi_+,\frac{e}{r}\}$, by a similar procedure we find
\begin{equation}\label{SL2R variation vanish}
\hat \delta H_{\eta_-}=\hat \delta H_{\eta_0}=\hat \delta H_{\eta_+}=0\,.
\end{equation}
Hence, by the choice of reference $H_{\eta_{_{0,\pm}}}[\bar\Phi]=0$ for an arbitrary member $\bar\Phi$ of these solutions, one arrives at
\begin{equation}\label{SL2R vanish}
H_{\eta_-}=H_{\eta_0}=H_{\eta_+}=0 .
\end{equation}
This result is of course expected, as a consequence of invariance under the non-Abelian $SL(2,\mathbb{R})$ exact symmetry group. 
\paragraph{Electric charge.} {Choosing $\eta=\{0,1\}$ as generator of  the electric charge $Q$, the $\delta H_\eta$ as integration of \eqref{NHEKN-k} would reduce simply to the standard result
\begin{equation}
\delta Q\equiv \delta H_\eta=\frac{-1}{16\pi G}\oint_{\partial \Sigma} \epsilon_{\mu\nu\mu_1\mu_{2}}\, \delta (\sqrt{-g}\delta F^{\mu\nu})\,\mathrm{d}x^{\mu_1}\!\!\wedge \mathrm{d}x^{\mu_{2}}\,,
\end{equation}
For the specific solution on which we have focused, by inserting parametric variations into the equation above we find $\hat\delta Q=\frac{\delta q}{G}$. So, by integration over parameters, electric charge of this geometry would be $Q=\frac{q}{G}$. }

\paragraph{Entropy.} {As discussed in Ref. \cite{HSS:2013lna}, in the geometry which we have focused on, any surface of constant time and radius, dubbed as $\mathcal{H}$, is the bifurcation point of a Killing horizon. Denoting those constant time and constant radius by $t_{_\mathcal{H}}$ and $r_{_\mathcal{H}}$, the associated horizon Killing vector is explicitly \cite{HSS:2013lna}
\begin{equation}
\tilde \zeta_{_\mathcal{H}} =-\frac{t_{_\mathcal{H}}^2r_{_\mathcal{H}}^2-1}{2r_{_\mathcal{H}}}\xi_-+ t_{_\mathcal{H}}r_{_\mathcal{H}}\xi_0-r_{_\mathcal{H}}\xi_+-k\partial_\varphi\,.
\end{equation} 
There are an infinite number of surfaces $\mathcal{H}$, hence an infinite number of Killing vectors $\tilde \zeta_{_\mathcal{H}}$. {The surface gravity associated to the $\tilde \zeta_{_\mathcal{H}}$ can be found to be $\kappa=1$ for any chosen $\mathcal{H}$ \cite{CHSS:2015bca}. So, the normalized horizon Killing vectors would be $\zeta_{_\mathcal{H}}=2\pi \tilde \zeta_{_\mathcal{H}}$.} Although $\zeta_{_\mathcal{H}}$ are Killings, they are not exact symmetries because $\delta _{\zeta_{_\mathcal{H}}}A=\frac{-2\pi er_{_\mathcal{H}}}{r^2}\mrd r\neq 0$. Instead, $\{\zeta_{_\mathcal{H}},\frac{-2\pi er_{_\mathcal{H}}}{r}+\text{const.}\}$ are exact symmetries.  The constant can be fixed by the integrability condition to be $2\pi e$ (see Appendix \ref{appen-B-2}), \emph{i.e.}
\begin{equation}\label{NHE KN eta H}
\eta_{_\mathcal{H}}=\{\zeta_{_\mathcal{H}},2\pi e(\frac{r-r_{_\mathcal{H}}}{r})\}.
\end{equation}
Note in particular that, similarly to $\zeta_{_\mathcal{H}}$, the exact symmetry generator $\eta_{_\mathcal{H}}$ also vanishes at the ``bifurcation surface'' $t=t_{_\mathcal{H}}, r=r_{_\mathcal{H}}$.   Inserting $\eta_{_\mathcal{H}}$ and the parametric variations in \eqref{NHEKN-k}, we obtain
\begin{equation}
\hat{\delta}H_{\eta_{_\mathcal{H}}}=\frac{4\pi a}{G}\delta a+\frac{2\pi q}{G}\delta q\,.
\end{equation}
Integrating over the solution phase space, the entropy is found as
\begin{equation}
S\equiv H_{\eta_{_\mathcal{H}}}=\frac{\pi(2a^2+q^2)}{G}\,.
\end{equation} 
}

Some comments and remarks are in order:
\begin{enumerate}
\item This geometry is \emph{not} a black hole, as it does not have an event horizon. Nonetheless, our formulation works for the calculation of conserved charges in this geometry.
\item The asymptotics of this geometry is not  flat or AdS. Our formulation  works for solutions with more general asymptotics.
\item According to the CPSM, the conserved charges of Killing vectors are independent of the chosen smooth and closed surface $\partial \Sigma$. This example exhibits that a similar feature extends to the case in the presence of matter fields.
\item The charges do not depend on the chosen $\partial \Sigma$, and one may in particular choose $\partial\Sigma$ to be the bifurcate Killing horizon surface for black hole solutions. Therefore, our formulation shows that all conserved charges, and hence the black hole thermodynamical quantities, are encoded in the near horizon geometry.  
\item It is worth mentioning that the ADM formalism \cite{Arnowitt:1959ah,Arnowitt:1960es,Arnowitt:1962hi} cannot be used to derive the angular momentum of this near horizon geometry (which is not asymptotic flat).  Also using the Komar integral \cite{Komar:1958wp} leads to incorrect angular momentum. This example shows one of the advantages of our proposed method over  these well-known methods. 
\end{enumerate}

\section{Black hole entropy and the first law of thermodynamics}\label{sec-5}

In previous sections we provided a formulation for computing conserved charge variations over the solution phase space. In this section we focus in particular on the notion of the entropy as a Hamiltonian generator associated with an exact symmetry. In this way we provide a generalization and extension of Wald's seminal result \cite{Wald:1993nt}, especially to the cases involving gauge fields. {Moreover, we also work out (or prove) the first law of thermodynamics.} This part is a variation as well as a generalization and extension of the Iyer-Wald formulation \cite{Iyer:1994ys}.

Although our formulation, as explicitly demonstrated in the two examples of the previous section, applies to more general cases, to illustrate explicitly how it works in the most familiar and simple examples, let us focus on black hole solutions to $d$-dimensional generally covariant gravity theory which includes some gauge fields $A^a$. We assume these black holes to be stationary and admit a nondegenerate Killing horizon (while neither of these conditions is crucial for our formulation to work). Let us denote the timelike Killing vector of the stationary black hole by $\partial_t$ and its possible other axial $U(1)$ isometries by $\partial_{\varphi_i}$. One can choose the coordinates $\{t,\varphi^i\}$ such that the charges (Hamiltonian generators) to these isometries are integrable and, in particular, assume that
\begin{equation}
M=H_{\partial_t}\,,\qquad J_i=-H_{\partial_{\varphi^i}}\,,
\end{equation}
where $M$ and $J_i$ denote the mass and angular momenta. In these coordinates, the horizon Killing vector would be
\begin{equation}\label{hat zeta H}
\tilde\zeta_{_\mathrm{H}}=\partial_t+\Omega^i_{_\mathrm H}\partial_{\varphi^i}\,,
\end{equation}
where the $\Omega^i_{_\mathrm{H}}$ are angular velocities of the horizon. Denoting the surface gravity of the black hole on its bifurcate horizon by $\kappa$, motivated by the integrability discussions of the example discussed in previous section, we define
\begin{equation}\label{zeta H}
\zeta_{_\mathrm{H}}=\frac{2\pi}{\kappa}\tilde\zeta_{_\mathrm{H}}= \frac{2\pi}{\kappa}(\partial_t+\Omega^i_{_\mathrm H}\partial_{\varphi^i})\,.
\end{equation}
Note that $\Omega_{_\mathrm{H}}^i$, although constants over spacetime, are functions on the solution phase space, and hence it is not immediate that the charge associated with $\zeta_{_\mathrm{H}}$ is integrable. We will comment on this point further in the next section.\footnote{Similarly, it is not trivial that the charge associated with exact symmetries $\partial_t$ or $\partial_{\varphi_i}$ is integrable. This point needs explicit examination.} Moreover,  the reason  we consider the specific combination of Killings $\zeta_{_\mathrm{H}}$ will become apparent below.

\subsection{Entropy as a Hamiltonian generator and in cases involving gauge fields}
In  Refs. \cite{Wald:1993nt,Iyer:1994ys} Iyer and Wald  proposed that the entropy of these black holes is a conserved charge associated with the horizon generating Killing vector. In their analysis what was important was the vanishing of the vector at the codimension-2 bifurcation surface and the ``normalization'' factor $(2\pi)/\kappa$ was included to get the standard expected results for the entropy and/or for the first law. Starting from $\zeta_{_\mathrm{H}}$ \eqref{zeta H},  they provide a very elegant proof/derivation of  the first law of black hole thermodynamics for charge perturbations associated with any perturbations of dynamical/physical $\delta\Phi$ which satisfy linearized field equations.  This proof works very well in  the absence of gauge fields. In the presence of gauge fields, the proof misses producing the  $\Phi^a_{_\mathrm{H}}\delta Q_a$ term. It is basically because in 
$\zeta_{_\mathrm{H}}$ there is no trace of the gauge fields. Of course, there have been some analyses, \eg Refs. \cite{Sudarsky:1992ty,Sudarsky:1993kh, McCormick:2013nkb} and most notably recently in Ref. \cite{Prabhu:2015vua},  to include the effects of the gauge fields in this derivation to include the effects of the gauge field. Below, we give a ``natural'' extension of the Iyer-Wald formulation for the latter. 

Our main observation in this regard is to replace the notion of ``invariance up to internal gauge transformations'' \cite{Prabhu:2015vua} with the notion of ``exact symmetries,'' requiring one to generalize the notion of Killing $\zeta_{_\mathrm{H}}$ to  $\eta_{_\mathrm{H}}=(\zeta_{_\mathrm{H}}, \lambda^a_{\zeta_{_\mathrm{H}}})$ such that $\eta_{_\mathrm{H}}$ is an exact symmetry. This formulation, among other things, extends the notion of Wald (or Iyer-Wald) entropy as a conserved charge defined through an integration over the bifurcation surface. As we discussed, dealing with an exact symmetry, the integration surface could now be chosen to be any codimension-2, smooth, closed  spacelike hypersurface  $\partial \Sigma$. This will then also allow us to present a ``generalized'' derivation for the first law which automatically includes the gauge fields.

In the presence of gauge fields, one may show that the horizon generating Killing vector field is either \emph{not an exact symmetry} or is \emph{not satisfying the integrability condition} \eqref{integrability cond para}.
For the former, as mentioned above, one can always locally construct an exact symmetry based on $\zeta_{_\mathrm{H}}$ by the addition of appropriate gauge symmetry transformation. This is due to the fact that, since $\zeta_{_\mathrm{H}}$ is a symmetry of the solution,
\be
{\cal L}_{\zeta_{_\mathrm{H}}}F^a=0,
\ee
where $F^a=\mrd A^a$. Using Cartan's identity ${\cal L}_\xi X=\xi\cdot \mrd X-\mrd(\xi\cdot X)$, we learn $\mrd(\zeta_{_\mathrm{H}}\cdot \mrd A^a)=0$ or $\zeta_{_\mathrm{H}}\cdot \mrd A^a=\mrd{\cal Z}_{_\mathrm{H}}^a$ locally, where ${\cal Z}_{_\mathrm{H}}^a$ is a scalar function.  Then, using once again the Cartan identity,  $\delta_{\zeta_{_\mathrm{H}}}A^a-\mrd(\zeta_{_\mathrm{H}}\cdot A^a)=\mrd{\cal Z}_{_\mathrm{H}}^a$, we learn that
\be
\delta_{\zeta_{_\mathrm{H}}}A^a=\mrd(\zeta_{_\mathrm{H}}\cdot A^a+{\cal Z}^a_{_\mathrm{H}})\equiv \mrd\lambda^a_{\zeta_{_\mathrm{H}}}.
\ee
That is, $\zeta_{_\mathrm{H}}$ is an isometry generator \emph{up to a gauge transformation} (see Ref. \cite{Prabhu:2015vua} for a recent discussion on this).
As a result, the 
\be
\eta_{_\mathrm{H}}=\{ \zeta_{_\mathrm{H}},-\lambda^a_{\zeta_{_\mathrm{H}}}+\text{const.}\}
\ee
 would be a generator of an exact symmetry. The ``const." can then be determined by the integrability condition over the solution phase space ${\cal F}_p$.

Although we have examples of solutions for which $\lambda^a_{\zeta_{_\mathrm{H}}}$ is not zero (\emph{cf.} the near horizon extremal Kerr-Newman geometry discussed in Sec. \ref{sec-4-2}) for  the known black hole solutions we have examined, the solution is written in a gauge where $\zeta_{_\mathrm{H}}$ is an exact symmetry, \ie $\delta_{\zeta_{_\mathrm{H}}}A^a=0$ and $\lambda^a_{\zeta_{_\mathrm{H}}}=0$. So, let us focus on these cases. One may then, using an analysis similar to those worked through in Appendix \ref{appen-B},  check that the \emph{integrability condition} \eqref{integrability cond para} fixes the constant part, leading to
\begin{equation}\label{eta-H-BH-entropy}
\eta_{_\mathrm{H}}=\{ \zeta_{_\mathrm{H}},-\dfrac{2\pi}{\kappa}\Phi^a_{_\mathrm{H}}\}\,,
\end{equation}
where $\Phi^a_{_\mathrm{H}}=\tilde\zeta_{_\mathrm{H}}\cdot A^a|_{\text{horizon}}$ are the horizon electric potentials. The details can be found in Appendix \ref{appen-B-3}.

The entropy for this family of solutions can be defined as follows.
\begin{center}
\emph{For stationary nonextremal black holes as solutions to covariant gravitational theories, the entropy $S$ can be defined as the Hamiltonian generator associated to the $\eta_{_\mathrm{H}}$ calculated on any smooth and closed $d\!-\!2$-dimensional surface surrounding the black hole.} 
\end{center}
\noindent Although this definition reproduces similar results as the Iyer-Wald entropy, it differs from it  in three aspects:
\begin{enumerate}
\item Entropy is defined as Hamiltonian generator $H_{\eta_{_\mrH}}$, instead of the Noether-Wald charge.
\item The surface of integration is relaxed and need not be the horizon itself. In particular, note that the exact symmetry generator $\eta_{_\mathrm{H}}$ does not/need not vanish at the horizon bifurcation surface.
\item The generator contains a specific gauge transformation, when gauge fields are present. This extra term is fixed by the condition of $\eta_{_\mathrm{H}}$ being an exact symmetry and the integrability of the associated charge over the solution phase space ${\cal F}_p$.
 We note that, even for cases where $\zeta_{_{\rm H}}$ is an exact symmetry (like, e.g. the Kerr-Newman black hole), as pointed out above, the charge associated with  $\zeta_{_{\rm H}}$ (which produces  the Wald entropy in uncharged cases)  is \emph{not} integrable. To make it integrable one needs to add an appropriately chosen gauge transformation, as given in \eqref{eta-H-BH-entropy}. 
\end{enumerate}

\subsection{Revisiting derivation of the first law}

Given the above definition of the entropy, we revisit the Iyer-Wald proof of the first law, matching it with our definition of the entropy. Since many of the details of the computations are essentially a repetition of Iyer-Wald analysis \cite{Iyer:1994ys}, we do not present them and here we only highlight the modifications. Assume that a stationary non-extremal black hole is given as a solution to a covariant gravitational theory in $d$-dimensional spacetime. For the generator $\eta_{_\mathrm{H}}$, we have  $\delta_{\eta_{_\mathrm{H}}} \Phi=0$. So, the Lee-Wald symplectic current density\footnote{As already pointed out, dealing with exact symmetries, our analysis here goes through for any other (pre)symplectic current density as well. }
\begin{equation}
\boldsymbol{\omega}(\delta \Phi,\delta_{\eta_{_\mathrm{H}}}\Phi,\Phi)=\delta\mathbf{\Theta}(\delta_{ \eta_{_\mathrm H}}\Phi,\Phi)-\delta_{\eta_{_\mathrm{H}}}\mathbf{\Theta}(\delta\Phi,\Phi)
\end{equation}
which is linear in $\delta_{\eta_{_\mathrm{H}}} \Phi$ vanishes.  Integrating over an arbitrary $d\!-\!1$-dimensional spacelike surface with two closed and smooth boundaries $\partial_1\Sigma$ and $\partial_2\Sigma$ surrounding the black hole, 
\begin{align}
\int_\Sigma  \boldsymbol{\omega}(\delta \Phi,\delta_{\eta_{_\mathrm{H}}}\Phi,\Phi)&=\int_\Sigma \mrd \boldsymbol{k}_{\eta_{_\mathrm{H}}}(\delta \Phi,\Phi)\\
&=\oint_{\partial_2\Sigma}\boldsymbol{k}_{\eta_{_\mathrm{H}}}(\delta \Phi,\Phi)-\oint_{\partial_1\Sigma}\boldsymbol{k}_{\eta_{_\mathrm{H}}}(\delta \Phi,\Phi).
\end{align} 
Since  $\delta_{\eta_{_\mathrm{H}}} \Phi=0$,
\begin{equation}
\oint_{\partial_1\Sigma}\boldsymbol{k}_{\eta_{_\mathrm{H}}}(\delta \Phi,\Phi)=\oint_{\partial_2\Sigma}\boldsymbol{k}_{\eta_{_\mathrm{H}}}(\delta \Phi,\Phi)\,,
\end{equation}
and one may drop the index 1 or 2 on ${\partial_i\Sigma}$ and define the entropy perturbation as
\begin{equation}
\delta S=\oint_{\partial\Sigma}\boldsymbol{k}_{\eta_{_\mathrm{H}}}(\delta \Phi,\Phi)\,.
\end{equation}
The generator $\eta_{_\mathrm{H}}=\{\zeta_{_\mathrm{H}},-\frac{2\pi}{\kappa}\Phi^a_{_\mathrm{H}}\}$ can be decomposed as
\begin{equation}
\eta_{_\mathrm{H}}=\{\frac{2\pi}{\kappa}\partial_t,0\}+\{\frac{2\pi\Omega^i_{_\mathrm{H}}}{\kappa}\partial_{\varphi^i},0\}+\{0,-\frac{2\pi}{\kappa}\Phi^a_{_\mathrm{H}}\}\,.
\end{equation}
So, by the linearity of $\boldsymbol{k}$ in generators,
\begin{equation}
\delta S=\frac{2\pi}{\kappa}\oint_{\partial\Sigma}\boldsymbol{k}_{\partial_t}(\delta \Phi,\Phi)+\frac{2\pi\Omega^i_{_\mathrm{H}}}{\kappa}\oint_{\partial\Sigma}\boldsymbol{k}_{\partial_{\varphi^i}}(\delta \Phi,\Phi)-\frac{2\pi}{\kappa}\Phi^a_{_\mathrm{H}}\oint_{\partial\Sigma}\boldsymbol{k}_{_{\{0,1\}}}(\delta \Phi,\Phi)\,.
\end{equation}
Noticing that the generators for mass, angular momenta, and electric charge are $\{\partial_t,0\}$, $\{\partial_{\varphi^i},0\}$, and $\{0,1\}$ respectively (with an additional minus sign for angular momenta),  then
\begin{equation}
\delta S=\frac{2\pi}{\kappa}\delta M-\frac{2\pi}{\kappa}\Omega^i_{_\mathrm{H}}\delta J_i-\frac{2\pi}{\kappa}\Phi^a_{_\mathrm{H}}\delta Q\,.
\end{equation}
Finally, by rearrangement and the Hawking temperature $T_{_\mathrm{H}}=\frac{\kappa}{2\pi}$, the first law of black hole thermodynamics is proved:
\begin{equation}\label{the first law}
\delta M=T_{_\mathrm{H}}\delta S+\Omega^i_{_\mathrm{H}}\delta J_i+\Phi^a_{_\mathrm{H}}\delta Q_a.
\end{equation}
As some remarks, we emphasize that our proof and results are independent of $\partial\Sigma$. Moreover, for the above proof to go through,  the only condition on $\delta \Phi$ is to satisfy linearized equation of motion, as in the Iyer-Wald case \cite{Iyer:1994ys}. In particular, the parametric variations $\hat\delta \Phi$ can also be used. In this case one obtains the Bardeen-Carter-Hawking form of the first law \cite{Bardeen:1973gd} which in our notation is $\hat\delta M=T_{_\mathrm{H}}\hat\delta S+\Omega^i_{_\mathrm{H}}\hat\delta J_i+\Phi^a_{_\mathrm{H}}\hat\delta Q_a$.

\section{ Discussion and outlook}\label{sec-6}

In this paper we revisited the derivation of conserved charges in generally covariant theories with a particular emphasis on the cases which also involve (internal) gauge symmetries. We based our analysis on the covariant phase space method and the notion of exact symmetries. We stressed the importance of the solution phase space ${\cal F}_p$ and parametric variations $\hat\delta\Phi$. The solution phase space is composed of a set of solutions to the classical field equations which are parametrized by some parameters $p_\alpha$, $\Phi(x^\mu;p_\alpha)$.\footnote{{Note that both $\Phi$ and $\hat\delta\Phi$ are defined up to pure gauge transformations.}} The symplectic structure of the phase space is the Lee-Wald symplectic structure possibly plus (appropriate) boundary $\mathbf{Y}$-terms. The tangent space of this phase space consists of  parametric variations $\hat\delta \Phi$; $\hat\delta \Phi$ may be viewed as one-forms in this tangent space. This phase space is expected to be included in any presumed bigger phase space, \emph{e.g.} those introduced and discussed in Refs. \cite{Lee:1990gr,Wald:1999wa,Barnich:2007bf}.

We then used the covariant phase space method applied to the solution phase space ${\cal F}_p$  for calculating conserved charges associated with exact symmetries.  For any given exact symmetry generator $\eta$ and parametric variation $\hat\delta \Phi$, one can associate conserved charge variations $\hat \delta H_\eta$ over the ${\cal F}_p$. $\hat \delta H_\eta$ is obtained from integration of $\boldsymbol{k}_\eta(\hat{\delta}\Phi, \Phi)$ which a surface term  read from the symplectic current density $\boldsymbol{\omega}$ and the latter is computed for any given generally covariant theory. The integration can be performed on any smooth and closed $d\!-\!2$-dimensional surface. 

 $\hat \delta H_\eta$ will lead to the conserved charge (Hamiltonian generator) $ H_\eta$ over ${\cal F}_p$, iff $\hat \delta H_\eta$ is an exact one-form on the tangent space of ${\cal F}_p$. The latter, is nothing but the integrability condition for the charge, which we worked out in \eqref{integrability cond para}. Once integrability is established, one may integrate $\hat \delta H_\eta$ over a path on the solution phase space to obtain the charges  $H_\eta(\Phi)$. Integrability then implies that the charges are independent of the path. In this formulation the charges are then specified up to an integration constant which is fixed upon an appropriate choice for the reference point (reference solution).

Here we highlight some notable features of our formulation for defining conserved charges associated with exact symmetries and some open problems in this regard.

\paragraph{Covariance of the charges.} Since our formulation is based on the covariant phase space method, our method provides a covariant definition of the charges, and relaxes the dependence on any specific integration surface (horizon or asymptotic region). 
\paragraph{Relaxing the dependence of charges on asymptotic behavior of the solution.}  Our charge variations are defined on the solution phase space ${\cal F}_p$, and the charges are obtained by an integration over the a path in ${\cal F}_p$. The corresponding integration constant is fixed by a reference solution there, and not the asymptotic behavior of the geometry, as is usually prescribed in ADM formulation or its extensions and variants to asymptotically AdS spaces.
\paragraph{Black hole entropy as a Hamiltonian generator.} As an immediate application of the formulation, we revisited the Iyer-Wald definition of entropy for black holes and discussed that it conveniently extends the (Iyer-)Wald  definition in two ways: (1) The entropy may now be defined as a conserved charge associated with an exact symmetry and hence can be obtained by integration over a generic codimension-2 surface $\partial\Sigma$ (and not just the horizon bifurcation surface). (2) It is not a Noether-Wald charge associated with the horizon generating Killing vector field but  a  Hamiltonian generator associated with an exact symmetry. The latter brings in the effects of the gauge fields.  Consequently, we observe that the entropy is not necessarily only a property associated with the geometry and its horizon, but it is a property of the solution.
\paragraph{Doing away with the horizon?} In the standard derivation/definition of the entropy using Wald's recipe, one is prescribed to integrate the charge associated with the horizon generating Killing vector field $\zeta_{_\mathrm{H}}$, which is defined by an integration at the bifurcation surface of the horizon. In this formulation to remove the ambiguities of the Noether-Wald charges one crucially uses two properties of $\zeta_{_\mathrm{H}}$  \cite{Iyer:1994ys,Wald:1993nt}: that $\zeta_{_\mathrm{H}}$ vanishes \emph{at the bifurcation surface} and is null \emph{at the horizon}. In our formulation, as already stressed, we are defining all charges, including the entropy, as a Hamiltonian generator associated with an exact symmetry. In particular,  we define the entropy by integration over any generic codimension two surface, where neither of these two properties holds. Despite the fact that the role of horizon and properties of $\zeta_{_\mathrm{H}}$ is ``weakened'' in our setting, in the definition of the entropy, we based the construction of the associated exact symmetry $\eta_{_\mathrm{H}}$  on the same Killing vector field $\zeta_{_\mathrm{H}}$. It is desirable to analyze this point further to check if even this much could be relaxed.

\paragraph{Symplectic exact symmetries vs. nonexact symmetries.} 
 In the end of Sec. \ref{sec-2} we discussed two interesting cases for which the (pre)symplectic current density $\boldsymbol{\omega}$ vanishes on-shell [\emph{cf.} \eqref{symplectic-condition} and discussions around it]: the symplectic nonexact and exact symmetries. The two examples of the former were analyzed recently \cite{Compere:2015knw, CHSS:2015mza,CHSS:2015bca}. The analysis in these works reveals an interesting feature: the charges associated with the nonexact symmetries and those associated with exact symmetries are two distinct sets and importantly commute with each other.\footnote{Although the simple Lie bracket of the corresponding symmetry generators may not vanish, once we recall field dependence of the generators and use the ``adjusted bracket'' \cite{Compere:2015knw}, the adjusted bracket of nonexact and exact symmetries vanishes.}  It seems that this property is independent of specific features of these  particular examples and is more general. It is interesting to establish this in a general setting.

Moreover, this also reveals that for the cases with nonexact and exact symmetries one can construct a  ``bigger phase space,'' ${\cal F}_{{\boldsymbol{\omega}}}$, of which the tangent space at each point $\Phi$ contains the set of perturbations $\delta_{\boldsymbol{\omega}}\Phi$ which is the union of the set of nonexact symmetries $\delta_\chi\Phi$ and exact symmetries $\hat\delta\Phi$. Since these two sets have no overlap, and if the above statement/expectation that the corresponding charges commute holds, then ${\cal F}_{{\boldsymbol{\omega}}}={\cal F}_{\chi} \otimes {\cal F}_p$, where ${\cal F}_{\chi}$ denotes the phase space associated with nonexact symmetries. One of the consequences of the latter is that ${\cal F}_{{\boldsymbol{\omega}}}$  denotes the the union of the coadjoint orbits of the nonexact symmetry algebra and that these coadjoint orbits are labelled by the charges associated with exact symmetries. The prime example of this is the set of locally AdS$_3$ geometries discussed in Ref. \cite{Compere:2015knw}. It is desirable to establish this in the case of other examples and hopefully in an example-independent way.

\paragraph{Acknowledgements.} We would like to thank Quantum Gravity Group at IPM.  We also thank Glenn Barnich, Geoffrey Comp\`ere, Reza Javadi-Nejad and especially Ali Seraj for helpful comments and discussions. This work has been supported by the \emph{Allameh Tabatabaii} Prize Grant of \emph{National Elites Foundation} of Iran and the \emph{Saramadan grant} of the Iranian vice presidency in science and technology. This work is also supported in part by the ICTP network scheme, NET-68.

\appendix

\addtocontents{toc}{\protect\setcounter{tocdepth}{0}}

\section{Noether-Wald charges in presence of internal gauge symmetries}\label{appen-A}

The variation of Lagrangian under transformations generated by $\epsilon=\{\xi,\lambda^a\}$ is
\begin{equation}\label{lagrangian deviation}
\delta _\epsilon \mathbf{L}=\mathbf{E}_{\Phi} \delta _\epsilon\Phi +\mathrm{d}\mathbf{\Theta} (\delta_\epsilon\Phi,\Phi)\,,
\end{equation}
where summation on different dynamical fields should be understood in $\mathbf{E}_\Phi \delta _\epsilon\Phi$ and 
\be
\delta_\epsilon \Phi=\mathscr{L}_\xi\Phi+\delta_{\lambda^a}A^a. 
\ee
The field equations for $\Phi$ are $\mathbf{E}_\Phi= 0$. According to the identity $\delta _\xi \mathbf{L}= \xi  \cdot  \mathrm{d} \mathbf{L} +\mathrm{d} (\xi  \cdot \mathbf{L})$ and noting that $\delta_{\lambda^a}\mathbf{L}=0$ and $\mathrm{d} \mathbf{L}=0$, we can replace the lhs of \eqref{lagrangian deviation} by $\mathrm{d} (\xi  \cdot \mathbf{L})$, so
\begin{equation}
\mathrm{d}\mathbf{\Theta}(\delta_\epsilon\Phi,\Phi)-\mathrm{d} (\xi  \cdot \mathbf{L}) \approx 0 \,,
\end{equation}
where $\approx$ denotes on-shell equality. We can now introduce a Noether $(d\!-\!1)$-form  current $\mathbf{J}_\epsilon$ as
\begin{equation}\label{Noether Wald current}
{\mathbf{J}_\epsilon \equiv \mathbf{\Theta}(\delta_\epsilon\Phi,\Phi)-\xi \! \cdot \! \mathbf{L}}\,.
\end{equation}
One may check that  $\mathrm{d} \mathbf{J}_\epsilon\approx 0$. According to Poincar\'e's lemma, since $\mathbf{J}_\epsilon$ is closed, it would be locally exact on-shell, and can be written as
\begin{equation}\label{Q-form}
{\mathbf{J}_\epsilon\approx\mathrm{d} \mathrm{\mathbf{Q}}}_\epsilon\,,
\end{equation}
where $\mathrm{\mathbf{Q}}_\epsilon$ is a $(d\!-\!2)$-form, the \emph{Noether-Wald charge density}. By variation of \eqref{Noether Wald current}, we have
\begin{equation}\label{H v Q proof 2}
\delta{\mathbf{J}_\epsilon = \delta\mathbf{\Theta}( \delta_\epsilon\Phi,\Phi)-\xi \! \cdot \! \delta\mathbf{L}}\,.
\end{equation}
In the standard CPSM, it is assumed that $\delta \epsilon=0$, so $\delta$ passes through it. Substituting $\delta \mathbf{L}$ in the last term in \eqref{H v Q proof 2} by \eqref{delta L Theta},
\begin{align}
\delta\mathbf{J}_\epsilon &\approx \delta\mathbf{\Theta}( \delta_\epsilon\Phi,\Phi)-\xi \! \cdot \! \mrd \mathbf{\Theta}(\delta \Phi,\Phi)\cr
&=\delta\mathbf{\Theta}( \delta_\epsilon\Phi,\Phi)-\mathscr{L}_\xi \mathbf{\Theta}(\delta \Phi,\Phi)+\mrd \big(\xi \cdot \mathbf{\Theta}(\delta \Phi,\Phi)\big)\,.
\end{align}
By rearrangement, and by $\delta\mathbf{J}_\epsilon= \delta \mrd \mathbf{Q}_\epsilon=\mrd \delta \mathbf{Q}_\epsilon$ where the last equality is a result of linearized field equations, we find
\begin{equation}\label{H v Q proof 1}
\delta\mathbf{\Theta}( \delta_\epsilon\Phi,\Phi)-\mathscr{L}_\xi \mathbf{\Theta}(\delta \Phi,\Phi)\approx \mrd \big(\delta \mathbf{Q}_\epsilon-\xi \cdot \mathbf{\Theta}(\delta \Phi,\Phi)\big)\,.
\end{equation}
For the theories under considerations, $\mathbf{\Theta}(\delta\Phi,\Phi)$ is composed of some gauge invariant quantities (\emph{e.g.} the field strengths $F^a=\mrd A^a$) and $\delta \Phi$. On the other hand, by the $\delta_{\lambda^a}\delta A^b=0$, the $\delta_{\lambda^a}\delta \Phi=0$ can be deduced. So, $\delta_{\lambda^a}\mathbf{\Theta}(\delta\Phi,\Phi)=0$. By subtraction of this vanishing term from the lhs of \eqref{H v Q proof 1}, we end with
\begin{equation}\label{H v Q proof 3}
\delta\mathbf{\Theta}( \delta_\epsilon\Phi,\Phi)-\delta_\epsilon\mathbf{\Theta}(\delta \Phi,\Phi)\approx \mrd \big(\delta \mathbf{Q}_\epsilon-\xi \cdot \mathbf{\Theta}(\delta \Phi,\Phi)\big)\,.
\end{equation} 
Notice that this result is correct, irrespective of any chosen $\mathbf{Y}$ ambiguity. It is because in \eqref{Noether Wald current} any $\mathbf{\Theta}\to \mathbf{\Theta}+\mrd \mathbf Y$ would also yield the  $ \mathbf{Q}_\epsilon\to \mathbf{Q}_\epsilon+\mathbf{Y}(\delta_\epsilon\Phi)$ through \eqref{H v Q proof 2}, keeping the \eqref{H v Q proof 3} intact. Comparing \eqref{H v Q proof 3} with \eqref{LW dk}, the explicit general formula for $\boldsymbol{k}_\epsilon(\delta\Phi,\Phi)$ can be read as
\begin{equation}
\boldsymbol{k}_\epsilon(\delta\Phi,\Phi)=\delta \mathbf{Q}_\epsilon-\xi \cdot \mathbf{\Theta}(\delta \Phi,\Phi)\,.
\end{equation}

\paragraph{Integrability.}

The integrability condition is explicitly 
\begin{equation*}
(\delta_1\delta_2-\delta_2\delta_1)H_\epsilon (\Phi)=0\,,\qquad \forall \Phi\in{\cal F},\ \delta\Phi\in T_{\cal F}\,.
\end{equation*}
We can use \eqref{k_xi general} to rewrite it as
\begin{align}
(\delta_1\delta_2 -\delta_2\delta_1)H_\epsilon &\approx \oint (\delta_1\delta_2 -\delta_2\delta_1)\mathbf{Q}_\epsilon -\oint \Big(\delta_1(\xi\cdot\mathbf{\Theta}(\delta_2\Phi,\Phi))-\delta_2(\xi\cdot\mathbf{\Theta}(\delta_1\Phi,\Phi))\Big)\label{integ cond proof 1}\\
&=-\oint \Big(\xi\cdot\delta_1\mathbf{\Theta}(\delta_2\Phi,\Phi)-\xi\cdot\delta_2\mathbf{\Theta}(\delta_1\Phi,\Phi)\Big)\label{integ cond proof 2}\\
&=-\oint \Big(\xi\cdot\big(\delta_1\mathbf{\Theta}(\delta_2\Phi,\Phi)-\delta_2\mathbf{\Theta}(\delta_1\Phi,\Phi)\big)\Big)\\
&=-\oint \xi\cdot \boldsymbol{\omega}(\delta _1\Phi,\delta_2\Phi,\Phi)\,.\label{integ cond proof}
\end{align}
Note that $\delta \mathbf{Q}_\epsilon$ is by definition integrable, \emph{i.e.} $(\delta_1\delta_2 -\delta_2\delta_1)\mathbf{Q}_\epsilon=0$, which is used in the equations above. Note also that  in deriving \eqref{integ cond proof 2} from \eqref{integ cond proof 1}, we have assumed $\delta \xi=0$. The vanishing of \eqref{integ cond proof} is the desired integrability condition \eqref{integrability-cond}. For the field-dependent variations, when $\delta\epsilon\neq 0$, we need to modify the above, as discussed in the Appendix of Ref. \cite{Compere:2015knw} and quoted in Sec. \ref{sec-2}.

\section{Integrability of $\eta_{_\mathrm{H}}$ for examples in Sec. \ref{sec-4}}\label{appen-B}

In this Appendix we investigate the integrability of  the exact symmetry generators associated with the entropy, $\eta_{_\mathrm{H}}$. In general $\eta_{_\mathrm{H}}$ is a linear combination of other exact symmetry generators which are integrable. Recalling the fact that the integrability condition \eqref{integrability cond para} is linear in the (exact) symmetry generators, the main point for the integrability condition to hold is the parameter dependence of the coefficients of linearity. Explicitly,  the integrability condition for $\eta=\{\zeta,\lambda\}$ is
 \begin{equation}\label{integ-cond-eta-H}
\oint_{\partial\Sigma} \Big(\zeta\cdot \boldsymbol{\omega}(\hat\delta_1\Phi,\hat\delta_2\Phi,\Phi)+\boldsymbol{k}_{\hat\delta_1\eta}(\hat\delta_2\Phi,\Phi) -\boldsymbol{k}_{\hat\delta_2\eta}(\hat\delta_1\Phi,\Phi)\Big)\approx0\,.
\end{equation}
If $\zeta$ is a linear combination of parameter-independent Killing vectors (exact symmetries) of which the charges associated are integrable, one may then drop the first term. For such cases the integrability condition 
reduces to 
\begin{equation}\label{integ-cond-eta-H-2}
\oint_{\partial\Sigma} \Big(\boldsymbol{k}_{\hat\delta_1\eta}(\hat\delta_2\Phi,\Phi) -\boldsymbol{k}_{\hat\delta_2\eta}(\hat\delta_1\Phi,\Phi)\Big)=\hat \delta_2 H_{\hat\delta_1\eta}-\hat\delta_1 H_{\hat\delta_2\eta}\approx0\,.
\end{equation}
In what follows for concreteness we will focus on the two examples discussed in Sec. \ref{sec-4}.

\subsection{Integrability $\eta_{_\mathrm{H}}$ for Kerr-AdS black hole}\label{appen-B-1}

As discussed in Sec. \ref{sec-4-1}, the charges associated with the two Killing vectors of the Killing vectors (exact symmetries) of the Kerr-AdS$_4$ geometry, $\partial_t, \partial_\varphi$ are integrable, as may be readily checked from \eqref{integ-cond-eta-H}. Let us start with a general linear combination of the two,
\be
\zeta=\mathrm{A}\partial_t+\mathrm{B}\partial_\varphi\,
\ee
where $\mathrm{A}$ and $\mathrm{B}$ are some constants over spacetime but functions of the parameters of the Kerr-AdS solution, \emph{i.e.} $\mathrm A$=$\mathrm A(m,a)$, $\mathrm B$=$\mathrm B(m,a)$. Since the $T_{\mathcal{F}_p}$ for the Kerr-AdS$_4$ solution is two dimensional and spanned by $\frac{\partial\Phi}{\partial m}\delta m$ and $\frac{\partial\Phi}{\partial a}\delta a$, to check \eqref{integ-cond-eta-H-2}, it is enough to only consider $\hat\delta_1\Phi=\frac{\partial\Phi}{\partial m}\delta m$ and $\hat\delta_2\Phi=\frac{\partial\Phi}{\partial a}\delta a$, yielding
\begin{align}
0&=\hat \delta_2 H_{\hat\delta_1\zeta}-\hat\delta_1 H_{\hat\delta_2\zeta}=\hat \delta_2 H_{\frac{\partial\mathrm{A}}{\partial m}\delta m\,\partial_t+\frac{\partial\mathrm{B}}{\partial m}\delta m\,\partial_\varphi}-\hat \delta_1 H_{\frac{\partial\mathrm{A}}{\partial a}\delta a\,\partial_t+\frac{\partial\mathrm{B}}{\partial a}\delta a\,\partial_\varphi}\cr
&=(\frac{\partial\mathrm{A}}{\partial m}\delta m) \,\hat \delta_2 H_{\partial_t}+(\frac{\partial\mathrm{B}}{\partial m}\delta m) \,\hat \delta_2 H_{\partial_\varphi}-(\frac{\partial\mathrm{A}}{\partial a}\delta a) \,\hat \delta_1 H_{\partial_t}-(\frac{\partial\mathrm{B}}{\partial a}\delta a) \,\hat \delta_1 H_{\partial_\varphi}\cr
&=(\frac{\partial\mathrm{A}}{\partial m}\delta m)(\frac{\partial M}{\partial a}\delta a) -(\frac{\partial\mathrm{B}}{\partial m}\delta m) (\frac{\partial J}{\partial a}\delta a)-(\frac{\partial\mathrm{A}}{\partial a}\delta a)(\frac{\partial M}{\partial m}\delta m)+ (\frac{\partial\mathrm{B}}{\partial a}\delta a)(\frac{\partial J}{\partial m}\delta m)\,, \nonumber
\end{align}
where $M=\frac{m}{\Xi^2}$ and $J=\frac{ma}{\Xi^2}$. In the above we used the fact that the charges for the linear combination of exact symmetries are the same linear combination of the corresponding charges. The above is satisfied if
\begin{align}\label{Kerr AdS integ proof}
\frac{\partial\mathrm{A}}{\partial m}\frac{\partial M}{\partial a}-\frac{\partial\mathrm{A}}{\partial a}\frac{\partial M}{\partial m}=\frac{\partial\mathrm{B}}{\partial m}\frac{\partial J}{\partial a}- \frac{\partial\mathrm{B}}{\partial a}\frac{\partial J}{\partial m}\,.
\end{align}
The above may also be written in a way convenient  for further analysis below,
\be\label{A-B-bracket}
\{\!\!\{\mathrm{A},M\}\!\!\}-\{\!\!\{\mathrm{B},J\}\!\!\}=0,\qquad \{\!\!\{X,Y\}\!\!\}\equiv  \frac{\partial X}{\partial m}\frac{\partial Y}{\partial a}-\frac{\partial{X}}{\partial a}\frac{\partial Y}{\partial m}.
\ee
The bracket is a standard  bracket on the space spanned by parameters $m,a$. {We should stress that $\{\!\!\{\ ,\ \}\!\!\}$ is not the bracket induced by the Lee-Wald symplectic structure on ${\cal F}_p$; we have introduced it for convenience in performing the computations.} 

Now, our task is to find coefficients $\mathrm{A}$ and $\mathrm{B}$ for which  Eq. \eqref{A-B-bracket} is satisfied. First of all, one may check by a direct straightforward and a bit lengthy  computation that
\be\label{A0-B0}
\mathrm{A}_0=\frac{2\pi}{\kappa},\qquad \mathrm{B}_0=\frac{2\pi \Omega_{\rm H}}{\kappa},
\ee
where the values for $\kappa$ and $\Omega_{{\rm H}}$ given in Sec. \ref{sec-4-1} are a solution to \eqref{A-B-bracket}. To this end one may use the fact the horizon radius $r_{_{\rm H}}$ is an implicit function of $m,a$, \emph{cf}. Eq. \eqref{r-H-Kerr-AdS}. One may then show that a general solution to \eqref{A-B-bracket} is of the form 
\be\label{general-A-B}
\mathrm{A}=f(M)+{D}(m,a)\mathrm{A}_0+C(JM) J,\qquad \mathrm{B}=g(J)+D(m,a)\mathrm{B}_0-C(JM) M,
\ee
where $f,g, C$ are arbitrary functions of the their argument, and $D$ is a function satisfying $\{\!\!\{D,M\}\!\!\}=\{\!\!\{D,J\}\!\!\}\Omega_{\rm H}$. We also note that, written in terms of the bracket on $(m,a)$ space, the integrability condition \eqref{A-B-bracket} remains invariant under a ``canonical'' transformation 
$m\to \tilde m,\ a\to \tilde a$ for which $\{\!\!\{\tilde m,\tilde a\}\!\!\}=1$.

\subsection{Integrability $\eta_{_\mathrm{H}}$ for near horizon extremal Kerr-Newman geometry}\label{appen-B-2}

Here, we analyze integrability of $\eta_{_\mathrm{H}}$ for the near horizon extremal Kerr-Newman geometry discussed in Sec. \ref{sec-4-2}. Let us consider the exact symmetry generator $\eta=\{\zeta, \lambda\}$ in which 
\be
\zeta=\mathrm{A}^-\xi_-+\mathrm{A}^0\xi_0+\mathrm{A}^+\xi_++\mathrm{B}\partial_\varphi, \qquad \lambda=\frac{e\mathrm A^+}{r}+\mathrm{C}.
\ee 
The coefficients $\mathrm{A}$, $\mathrm{B}$, and $\mathrm{C}$ can be functions of the parameters $a$ and $q$, but constant over spacetime. One can then decompose $\eta$ as
\begin{equation}
\eta=\{\mathrm{A}^-\xi_-,0\}+\{\mathrm{A}^0\xi_0,0\}+\{\mathrm{A}^+\xi_+,\frac{e\mathrm A^+}{r}\}+\{\mathrm{B}\partial_\varphi,\mathrm{C}\}\,.
\end{equation}
Since the charges associated with the first three terms in the above decomposition for $\eta$ are integrable (\emph{cf}. the discussions of Sec. \ref{sec-4-2}), and since the $SL(2,\mathbb{R})$ charges and their  variations vanish over the whole solution phase space \eqref{SL2R variation vanish}, the integrability condition should only be checked for the last term in $\eta$, which is
 \begin{equation}\label{NHE KN integ proof 2}
\oint_{\partial\Sigma} \Big((\mathrm{B}\partial_\varphi)\cdot \boldsymbol{\omega}(\hat\delta_1\Phi,\hat\delta_2\Phi,\Phi)+\boldsymbol{k}_{\{\hat\delta_1\mathrm{B}\,\partial_\varphi,\hat\delta_1\mathrm{C}\}}(\hat\delta_2\Phi,\Phi) -\boldsymbol{k}_{\{\hat\delta_2\mathrm{B}\,\partial_\varphi,\hat\delta_2\mathrm{C}\}}(\hat\delta_1\Phi,\Phi)\Big)\approx0\,.
\end{equation}

The first term in the integral vanishes because of the integrability of the angular momentum and the integrability of $\eta$ then implies
\begin{equation}\label{NHE KN integ proof 3}
\hat\delta_1\mathrm{B}\,\hat\delta_2 H_{\partial_\varphi}+\hat \delta_1 \mathrm{C} \,\hat\delta_2 H_{\{0,1\}}-\hat\delta_2\mathrm{B}\,\hat\delta_1 H_{\partial_\varphi}-\hat \delta_2 \mathrm{C} \,\hat\delta_1 H_{\{0,1\}}\approx0\,.
\end{equation}
The $T_{\mathcal{F}_p}$ for the solution which we have focused on is spanned by $\frac{\partial\Phi}{\partial a}\delta a$ and $\frac{\partial\Phi}{\partial q}\delta q$. So, in order to check \eqref{NHE KN integ proof 3}, putting $\hat\delta_1\Phi=\frac{\partial\Phi}{\partial a}\delta a$ and $\hat\delta_2\Phi=\frac{\partial\Phi}{\partial q}\delta q$ would be enough, yielding
\begin{equation}
(-\frac{\partial\mathrm{B}}{\partial a}\delta a)(\frac{\partial J}{\partial q}\delta q)+(\frac{\partial\mathrm{C}}{\partial a}\delta a)(\frac{\partial Q}{\partial q}\delta q)+(\frac{\partial\mathrm{B}}{\partial q}\delta q)(\frac{\partial J}{\partial a}\delta a)-(\frac{\partial\mathrm{C}}{\partial q}\delta q)(\frac{\partial Q}{\partial a}\delta a)= 0\,.\nonumber
\end{equation}
in which $J=\frac{\sqrt{a^2+q^2}}{G}a$ and $Q=\frac{q}{G}$. As in the previous case of Appendix \ref{appen-B-1}, the above may also be written in terms a bracket on the  space spanned by $(q,a)$, 
\be\label{B-C-bracket}
\{\!\!\{\mathrm{B},J\}\!\!\}-\{\!\!\{\mathrm{C},Q\}\!\!\}=0,\qquad \{\!\!\{X,Y\}\!\!\}\equiv  \frac{\partial X}{\partial a}\frac{\partial Y}{\partial q}-\frac{\partial{X}}{\partial q}\frac{\partial Y}{\partial a}.
\ee

It can  be easily checked that 
\be\label{B0-C0}
\mathrm{B}=\mathrm{B}_0=2\pi k,\qquad \mathrm{C}=\mathrm{C}_0=2\pi e,
\ee 
for values of $k$ and $e$ given in \eqref{NHE KN entities}, satisfy the equation above. Freedom in the choice of $\mathrm{A}$'s is basically the same as freedom in the choice of $\mathcal{H}$. Therefore, by the analysis above, we have shown that the generators $\eta_{_\mathcal{H}}$ introduced in \eqref{NHE KN eta H} have integrable charges.

As in the previous case, one may find more general solutions of \eqref{B-C-bracket}, based on \eqref{B0-C0}. The construction is very similar to the one given in \eqref{general-A-B} and we do not repeat it here.

\subsection{Integrability $\eta_{_\mathrm{H}}$ for a generic nonextreme charged black hole}\label{appen-B-3}

Having discussed the two examples in previous subsections, let us now analyze the integrability of $\eta_{_\mathrm{H}}$ for a generic black hole.
Consider a stationary black hole with axial $U(1)_{\varphi^i}$ isometries as a solution to a covariant gravitational theory with some Abelian gauge fields $A^a$. Assume that it is identified by the parameters $m$, $a_i$, and $q_a$ collectively denoted by $p_\alpha$. The generators for the mass $M$, angular momenta $J_i$, and electric charges $Q_a$ can be respectively chosen to be $\{\partial_t,0\}$, $\{\partial_{\varphi^i},0\}$ and $\{0,1^a\}$ where by $1^a$ we mean  $\lambda^a=1$ and $\lambda^b=0$ if $b\neq a$. 

Now, consider an exact symmetry as $\eta=\{\zeta,\mathrm{C}^a\}$ in which $\zeta=\mathrm{A}\partial_t+\mathrm{B}^i\partial_{\varphi^i}$.  The $\mathrm{A}$, $\mathrm{B}^i$, and $\mathrm{C}^a$ are constants over the spacetime, but functions of the parameters. Recalling the integrability of mass and angular momenta, one may use \eqref{integ-cond-eta-H} as the integrability condition for $\eta$, explicitly,
 \begin{equation}
\oint_{\partial\Sigma} \Big(\boldsymbol{k}_{\hat\delta_1\eta}(\hat\delta_2\Phi,\Phi) -\boldsymbol{k}_{\hat\delta_2\eta}(\hat\delta_1\Phi,\Phi)\Big)=\hat \delta_2 H_{\hat\delta_1\eta}-\hat\delta_1 H_{\hat\delta_2\eta}\approx0\,.
\end{equation}

By the decomposition 
\begin{equation}
\eta=\{\mathrm{A}\partial_t,0\}+\{\mathrm{B}^i\partial_{\varphi^i},0\}+\{0,\mathrm{C}^a\}
\end{equation}
then
\begin{equation}
\hat \delta_2 H_{\hat\delta_1 \mathrm{A} \partial_t+\hat \delta_1 \mathrm{B}^i\partial_{\varphi^i}+ \{0,\hat\delta_1 \mathrm{C}^a\}}-\hat \delta_1 H_{\hat\delta_2 \mathrm{A} \partial_t+\hat \delta_2 \mathrm{B}^i\partial_{\varphi^i}+ \{0,\hat\delta_2 \mathrm{C}^a\}}\approx0\,.
\end{equation}
The $T_{\mathcal{F}_p}$ for the solution is spanned by $\frac{\partial\Phi}{\partial p_\alpha}\delta p_\alpha$ (no summation over $\alpha$). Hence, the equation above should be satisfied for any $\hat\delta_1\Phi=\frac{\partial\Phi}{\partial p_\alpha}\delta p_\alpha$ and $\hat\delta_2\Phi=\frac{\partial\Phi}{\partial p_\beta}\delta p_\beta$ for all $\alpha$ and  $\beta$. It results in the system of equations (no summation over $\alpha$ and $\beta$)
\begin{equation}\label{entropy integrebility proof 1}
(\frac{\partial \mathrm{A}}{\partial p_\alpha}\delta p_\alpha)(\frac{\partial M}{\partial p_\beta}\delta p_\beta)-(\frac{\partial \mathrm{B}^i}{\partial p_\alpha}\delta p_\alpha)(\frac{\partial J_i}{\partial p_\beta}\delta p_\beta)+(\frac{\partial \mathrm{C}^a}{\partial p_\alpha}\delta p_\alpha)(\frac{\partial Q_a}{\partial p_\beta}\delta p_\beta)-[\alpha\leftrightarrow\beta]=0\,, \quad \forall \alpha,\beta\,.
\end{equation}
The above can be written  in a convenient way, if we use a collective notion for the charges and coefficients,
\be
{\cal C}^\rho \equiv (\mathrm{A}, \mathrm{B}^i, \mathrm{C}^a), \qquad {\cal Q}_\rho\equiv (M, -J_i, Q_a),
\ee  
in terms of which \eqref{entropy integrebility proof 1} take the form
\be\label{integ-general}
\frac{\partial {\cal C}^\rho}{\partial p_\alpha}\frac{\partial {\cal Q}_\rho}{\partial p_\beta}-\frac{\partial {\cal C}^\rho}{\partial p_\beta}\frac{\partial {\cal Q}_\rho}{\partial p_\alpha}=0
\ee
where the summation on $\rho$ is understood. 

One can check that for standard black holes such as the Kerr-Newman  and Myers-Perry black holes the above is satisfied. In particular, for the $4$-dimensional Kerr-Newman, 
$$\mathrm{A}=\frac{2\pi}{\kappa}, \quad\mathrm{B}=\frac{2\pi\Omega_{_\mathrm{H}}}{\kappa}, \quad \mathrm{C}=-\frac{2\pi\Phi_{_\mathrm{H}}}{\kappa},$$ 
and for Myers-Perry $d$-dimensional Myers-Perry,
$$\mathrm{A}=\frac{2\pi}{\kappa}, \quad\mathrm{B}^i=\frac{2\pi\Omega^i_{_\mathrm{H}}}{\kappa},
$$
solve the integrability condition \eqref{integ-general}.

\end{document}